%% file: main.tex
\documentclass[%
 aip,
 amsmath,amssymb,
 reprint,%
]{revtex4-1}

\usepackage{graphicx}
\usepackage{dcolumn}
\usepackage{bm}

\usepackage[utf8]{inputenc}
\usepackage[T1]{fontenc}
\usepackage{mathptmx}
\usepackage{mathrsfs}
\usepackage{braket}
\usepackage{dcolumn}
\usepackage{bm}
\usepackage{tikz}
\usepackage{nicefrac}
\usepackage{array}
\usepackage{graphics}
\usepackage{amssymb}
\usepackage{amsmath}
\usepackage{graphicx}
\usepackage{multirow}
\usepackage{color}
\usepackage{comment}
\usepackage[normalem]{ulem}
\usepackage{url}
\usepackage{mathrsfs}
\usepackage{tabularx}
\usepackage[caption=false]{subfig}
\usepackage{appendix}
\usepackage{float}
\usepackage[makeroom]{cancel}

\usepackage{csquotes}
\renewcommand{\b}[1]{\boldsymbol{#1}}
\newcommand{\nid}{\noindent}

\begin{document}

\preprint{AIP/123-QED}

\title{Dissipative Tunneling Rates through the Incorporation of First-Principles Electronic Friction in Instanton Rate Theory II: Benchmarks and Applications}

\author{Y. Litman}
\email{yairlitman@gmail.com}
\affiliation{MPI for the Structure and Dynamics of Matter, Luruper Chaussee 149, 22761 Hamburg, Germany}

\author{E. S. Pós }%
\affiliation{MPI for the Structure and Dynamics of Matter, Luruper Chaussee 149, 22761 Hamburg, Germany}

\author{C. L. Box}%
\affiliation{Department of Chemistry, University of Warwick, Coventry CV4 7AL, United Kingdom}

\author{R. Martinazzo}
\affiliation{Department of Chemistry, Università degli Studi di Milano, Via Golgi 19, 20133 Milano, Italy}

\author{R. J. Maurer}%
\affiliation{Department of Chemistry, University of Warwick, Coventry CV4 7AL, United Kingdom}

\author{M. Rossi}
\email{mariana.rossi@mpsd.mpg.de}
\affiliation{MPI for the Structure and Dynamics of Matter, Luruper Chaussee 149, 22761 Hamburg, Germany}

\date{\today}

\begin{abstract}
In part I, we presented the ring-polymer instanton with explicit friction  (RPI-EF) method and showed how  it can be connected to the \textit{ab initio} electronic friction formalism. This framework allows the calculation of tunneling reaction rates that incorporate the quantum nature of the nuclei and certain types of non-adiabatic effects (NAEs) present in metals. In this second part, we analyze the performance of RPI-EF on model potentials and apply it to realistic systems. For a 1D double-well model, we benchmark the method against numerically exact results obtained from 
multi-layer multi-configuration time-dependent Hartree calculations. We demonstrate that RPI-EF is accurate for medium and high friction strengths and less accurate for extremely low friction values. We also show quantitatively how the inclusion of NAEs lowers the cross-over temperature into the deep tunneling regime, reduces the tunneling rates, and in certain regimes,
steers the quantum dynamics by modifying the tunneling pathways. 
As a showcase of the efficiency of this method, we present a study of hydrogen and deuterium hopping between neighboring interstitial sites in selected bulk metals. The results show that multidimensional vibrational coupling and nuclear quantum effects have a larger impact than NAEs on the tunneling rates of diffusion in metals. 
Together with part I, these results advance the calculations of dissipative tunneling rates from first principles.  
\end{abstract}

\maketitle

\section{\label{sec:intro}Introduction}

The transport kinetics of  small molecules and atoms in metals
play an important role in technological applications in several areas, like fuel cells, batteries, and nuclear reactors, among others \cite{Mohtadi_NatREv_2016,Robertson_Metal_2015,Abderrazak__CorrREv_2018,Schlapbach_Nature_2001}.  In particular, accurate measurements of hydrogen diffusion  are still experimentally challenging since diffusion coefficients are very  sensitive to the microstructure of the material and the composition of the alloy~\cite{Volkl_book, JOTHI20152882, Fukai2005}.  As a consequence, reported values of diffusion constants by different groups can be scattered over a few orders of  magnitude.

First-principles calculations have the potential to provide an increased understanding of the transport dynamics of light particles in well-defined structures throughout a wide range of thermodynamic conditions (e.g., temperature and pressure), which can complement experiments and guide further developments.   
Most of the atomistic calculations performed to study nuclear dynamical processes rely on Newtonian dynamics on the ground-state adiabatic potential energy surface (PES), as given by the Born-Oppenheimer approximation (BOA). However, the coupling of the nuclear movement with electrons in the metal can easily induce electronic excitations, which represent a breakdown of the BOA and give rise to non-adiabatic effects (NAEs)~\cite{Nienhaus_SSR_2002,Wodtke_ChemSocRev_2016}. Furthermore, light particles such as hydrogen and deuterium can exhibit strong nuclear quantum effects (NQEs), which can either increase or decrease its mobility through metals~\cite{Kimizuka_PRB_2018,Kimizuka_PRM_2021}.
As a consequence, an understanding of the interplay of NQEs and  NAEs in the transport process of light atoms in realistic systems remains elusive. 

Several exact methods to simulate non-adiabatic quantum dynamics have been developed  in the last decades\cite{MCTDH_1,HEOM,Quapi}.
However, due to the unfavorable scaling of these  theoretical approaches with the number of degrees of freedom, efficient but accurate methods are required to capture NQEs and NAEs in high-dimensional systems or to be used with costly \textit{ab initio} potentials.  In part I of this paper, we presented the ring-polymer instanton with explicit friction (RPI-EF) theory for the calculation of dissipative thermal tunneling rates, 
  which has the potential to fulfill these requirements.
  Briefly, RPI-EF allows an easy and efficient incorporation
  of the electronic friction formalism, originally proposed by
 Hellsing and Persson \cite{Hellsing_1984} and 
 Head-Gordon and Tully\cite{Head_Gordon_1995}, into the 
  the semi-classical  ring-polymer  instanton (RPI) rate  theory\cite{Jor_review_2018}. 
  
  In this article, we show the merits and limitations of RPI-EF. We benchmark the accuracy of RPI-EF rate predictions by comparing them with numerically exact theories in model potentials. Subsequently, we examine the interplay of the reaction barrier height and friction strengths on the tunneling rate in 1D and 2D model potentials connected to a bath.  Finally, the new 
  approach is applied to hydrogen and deuterium hopping reactions in bulk transition metals, focusing on Pd, by employing Kohn-Sham density functional theory. These calculations allow a quantitative analysis of the impact of different effects on the rate constants, such as  the dimensionality of the system, NQEs, and NAEs. 
 
Part II of this paper is structured as follows: 
In Section~\ref{sec:Rate_Calculations} the methods employed to obtain the tunneling rates are briefly summarized and the simulation details for each one are specified. In Section \ref{sec:systems}, the model potentials and the systems treated from first-principles are described. Results of numerical simulations on low-dimensional models for position-independent and position-dependent friction tensors  are discussed in Section \ref{sec:res:DW} and \ref{sec:res:DDW}, respectively. Finally, the first-principles results for the hydrogen and deuterium hopping in metals are discussed in  Section \ref{sec:res:ab_initio}. Section \ref{sec:Conclusions} concludes by summarizing the main results and giving an outlook to future directions.

\section{Rate Calculation Methods}\label{sec:Rate_Calculations}
\subsection{Ring-polymer instanton (with explicit friction) calculations}

The RPI rate theory~\cite{JOR_JCP_2009, Arni_thesis} 
 is a semi-classical method that allows
 the calculation of tunneling rates.  RPI theory can be interpreted as the extension of Eyring transition state theory (TST) \cite{Eyring_1935_JCP} into the deep tunneling regime, since the rates are evaluated utilizing only a limited amount (often a single) special configuration, circumventing the necessity of real-time sampling. While in TST this special configuration is the first-order saddle-point connecting reactants and products on the potential energy surface (PES), in RPI rate theory  the special configurations are found at the first-order saddle-points of the extended space of the ring polymer (RP) potential. These trajectories in imaginary time are known as instanton trajectories. 
 The RPI rate expression is analogous to the one proposed by TST  and reads, 
\begin{equation}
\begin{split}
   \label{eq:Kinst1}
   k_\text{inst}(\beta) \propto
    e^{-U_P^\text{sys}(\b{\bar{q}})/\hbar} ,
       \end{split}
\end{equation}
\nid where
\begin{equation}
      \label{eq:U_P}
  U_P^\text{sys} (\bm{q}) =\\
 \sum_{k=1}^P 
 \sum_{i=1}^{3N} 
   m_i\frac{\omega_P^2}{2} (q_i^{(k)}-q_i^{(k+1)})^2  +  \\ 
     \sum_{k=1}^P V( q_1^{(k)},\dots q_{3N}^{(k)})
\end{equation}

\nid is the RP potential. In the previous equation, $q^{(k)}_i$ is the position of the $i$-th degree of freedom of the $k$-th bead of the ring polymer, $m_i$ is the mass of the  $i$-th degree of freedom, $N$ is the number of atoms, $P$ is the number of beads (replicas),
$\bm{q}$ is an abbreviated notation to represent all the degrees of freedom, $\b{\bar{q}}$ denotes the instanton geometry, 
and $\omega_P = (\beta_P \hbar)^{-1}$ with $\beta_P = \frac{1}{k_BPT}$.

In part I, we showed how RPI rate theory can be extended to compute tunneling rates for systems connected to a harmonic bath which simulates a dissipative environment. Irrespective of the dissipative mechanism, and assuming that the environment degrees of freedom adjust adiabatically to the system position, one can fully characterize the system-environment coupling by a position ($\bm{q}$) and frequency ($\lambda$) dependent friction tensor, $\tilde{\eta}(\bm{q}, \lambda)$. Moreover, when the frequency and position dependence are decoupled, we proved  that the RP potential that enters Eq. \ref{eq:Kinst1} is renormalized, adopting the form
\begin{equation}
\begin{split}\label{eq:eff-SD}
U^{\text{MF}}_P=&
U_P^\text{sys}
+\\&
\sum_{l=-P/2+1}^{P/2}\sum_{i=1}^{3N}
\frac{\omega_l}{2}
\bigg[
\sum_{k=1}^P
C_{lk}
\bigg(\int_{\bm{q}^\text{ref}}^{\bm{q}^{(k)}} 
\tilde{\eta}(\bm{q}',\omega_k)^{1/2}  \cdot d\bm{q}'\bigg)
\bigg]^2,
\end{split}
\end{equation}
\nid where  $\omega_l= 2\omega_P \sin(|l|\pi/P)$ are the free RP normal mode frequencies and $\bm{C}$ is the  transformation matrix between  the RP normal modes and Cartesian coordinates. In the limiting case of a position-independent friction, the previous expression can be  simplified to 
\begin{equation}
\begin{split}\label{eq:MF-SI}
U^{\text{MF}}_P =U_P^\text{sys} +
\sum_{l=-P/2+1}^{P/2}  \sum_{i=1}^{3N}  \frac{\tilde{\eta}(\omega_l){\omega_l}}{2}(Q_i^{(l)})^2,\\
\end{split}
\end{equation}
\nid where  $\bm{Q}^{(l)}$ represent the free RP normal mode coordinates.

The RPI calculations were performed using the i-PI \cite{i-pi2} code.
The forces and energies required by the algorithm were passed to i-PI from an external code through an interface based on internet sockets. The RPI-EF calculations required an extension of the i-PI communication protocol. We added to the existing communication of the energy, forces, and stresses from external codes to i-PI, the possibility to pass additional information as JSON-formatted strings. In this way, it is possible to communicate any type of data and, more importantly, when the data is numeric, it becomes available to be used by any implemented algorithm within i-PI. This enables the use of quantities beyond energies and forces, that change along the simulation, within different types of dynamics. 

The FHI-aims code\cite{FHI-AIMS} and an in-house python code were used in connection to i-PI for the DFT and model calculations, respectively. Transition-state geometries were obtained  either by the  string method\cite{Weinan_JCP_2007} combined with
the climbing image technique\cite{Graeme_JCP_2000} or using a minimum-mode-following algorithm \cite{Nichols_JCPS_1990}, as implemented in i-PI. The RPI calculations were initialized after finding the transition state, by stretching the transition-state geometry along the mode with imaginary frequency using a number of replicas between 10 and 16. 
Optimizations were started at a temperature of 10~K below the corresponding cross-over temperature, $T_c^\circ$.
After converging the instanton pathway for the first calculation, successive steps of temperature decrease and RP interpolation to increase the number of beads were performed until the target temperature was reached.
If required, further calculations with more beads were performed to guarantee that, in all cases and for all temperatures, the final rates were converged within a 10\% error~\cite{Beyer_PCL_2016,Litman_thesis}. See more details in section I of the Supplemental Information (SI).

\subsection{Multi-configuration time-dependent Hartree Calculations}

Numerically exact results for selected models in this paper were obtained with the 
multi-layer variant\cite{wan03:1289,man09:054109,ven11:044135} of the multi-configuration time-dependent Hartree method\cite{mey90:73,bec00:1,mey09:book} (ML-MCTDH), as implemented in the Heidelberg package\cite{mctdh:MLpackage}. MCTDH is a variational method that relies on optimal, time-dependent basis functions to alleviate the exponential scaling problem of standard methods based on direct expansions on time-independent basis. In the  MCTDH \emph{ansatz} the wave function  $\Ket{\Psi(t)}$ is expanded on orthogonal  configurations $\Ket{\Phi_J(t)}$, which in turn are products of `single particle' functions (SPFs),
\begin{equation}
\begin{split}\label{eq:MCDTDH1}
\Ket{\Psi(t)} = \sum_J A_J(t)\Ket{\Phi_J(t)} = \sum_{J}A_J(t) \prod_k \ket{\phi_{j_k}^{(k)}}(t),
\end{split}
\end{equation}
and both the expansion coefficients ($A_J$) and the SPFs ($\ket{\phi_{j}(t)}$) are variationally optimized. Here, $J=(j_1,..j_k,..j_F)$ is a multi-index, the index $k=1,2,..F$ runs over the single-particles, and $j_k=1,..n_k$ labels  the SPFs used for the $k^{\textrm{th}}$ mode. Conventional MCTDH uses single-particles for each degree of freedom or small groups thereof, and represents their SPFs by a direct expansion on a grid/basis-set (the so-called primitive grid) designed for the single particle at hand. This limits its capability of handling large systems. In contrast, in ML-MCTDH the single particles are high-dimensional modes whose SPFs are described by further MCTDH expansions employing lower dimensional SPFs. The procedure can be indefinitely iterated till reasonably small single-particles are defined that can be described on primitive grids. This recipe, similarly to tensor networks and matrix-product states\cite{RevModPhys2021}, endows the wave function with a hierarchical, flexible structure that allows the treatment of considerably larger systems. In particular, ML-MCTDH has been successfully applied to the calculation of thermal rate constants in condensed-phase problems\cite{ML-MCTDH_rates,Craig_JCP_2007} in the framework of the reactive flux-side approach,  
where $k(T)$ is given by the long-time limit of the equilibrium flux-side time-correlation function\cite{Miller1983,Miller1998}. 

In this work, we followed closely the original work by Wang and Thoss\cite{Wang2006a,Craig2007}, who introduced an importance sampling technique to recast the trace expression of the flux-side correlation function as an accessible ensemble-average over time-evolving wavepackects. In a nutshell, the evaluation of the rate constant is reduced to: i) a preparation step where the wavepackets are initialized  by combining (system) Boltzmannized flux eigenvectors with bath states drawn from the canonical ensemble of the uncoupled bath, ii) an equilibration step where imaginary-time dynamics introduces the correlations present in the coupled system, and iii) a propagation step where the real-time dynamics is followed up to the onset of the kinetic regime (the long-time limit alluded to above).  
Details about the calculations, including an overview of the flux-side approach, the Monte Carlo sampling, the tree structure of the ML-MCTDH wavefunction, the number of SPFs and the primitive grids used, are provided in section II of SI. Converged calculations were obtained with 50 bath modes and using 128-256 realizations for each value of $T$ and coupling strength. 

\section{Simulation details and parameters} \label{sec:systems}
\subsection{1D and 2D double well models} 

We analyze the performance of RPI-EF on a double-well model similar to the ones usually employed to study quantum dynamics in system-bath models~\cite{Topaler_JCP_1994,Craig_JCP_2005}. The potential energy surface of the system is given by 
\begin{equation}
\begin{split}
V_\text{DW}(q) = - \frac{1}{2} m \omega^{\ddag 2} (q-q_0)^2 + \frac{m^2\omega^{\ddag 4}}{16V_0}q^4,
\end{split}\label{eq:DW}
\end{equation}
where, unless otherwise specified, we set $m$ as the mass of atomic hydrogen and $\omega^\ddag=500$ cm$^{-1}$. 
The coupling between the system and the bath can be made position-dependent according to\cite{Straus_JCP_1993}
\begin{equation}
\begin{split}\label{eq:coupling_f}
f(q) = q[1 + \epsilon_1 \exp(-\Delta q^2/2) +\epsilon_2 \tanh(\Delta q)],
\end{split}
\end{equation}
where
$\Delta q =(q-q^\ddag)/\delta$, $\delta$ determines the length-scale of the nonlinear couplings, and $\epsilon_1$ and $\epsilon_2$ the magnitude of its symmetric and anti-symmetric components, respectively.
The calculations with position-independent friction were obtained by setting $\epsilon_1 =\epsilon_2= 0$.  Naturally, the position-dependent couplings of real systems do not generally adopt such simple forms. However, as it will be shown in Sec. \ref{sec:res:DDW}, this simplified form will prove sufficient to expose the importance of including this position-dependence in the tunneling rates.

We consider an Ohmic (linear) spectral density multiplied by an exponential cutoff, leading to
%

\begin{equation}
\begin{split}\label{eq:eta_laplace}
\tilde{\eta}_{il}(\bm{q}, \lambda) =  \frac{2}{\pi} \int_{0}^{+\infty}d\omega 
\tilde{\eta}_0
 \bigg(\frac{\partial f(\bm{q})_i}{\partial q_l}\bigg)^2
\frac{\lambda}{\omega^2+\lambda^2}e^{-\omega/\omega_c}
\end{split}
\end{equation}
\nid where $\tilde{\eta}_0$ is the static friction coefficient and, unless otherwise specified, we set $\omega_c=500$ cm$^{-1}$.

To illustrate the tensorial nature of the friction we also considered a two-dimensional double double-well (DDW) potential given by %
\begin{equation}
\begin{split}
V_\text{DDW}(q_1,q_2) = 
V_\text{DW}(q_1)  
+
V_\text{DW}(q_2)   + Cq_1q_2,
\end{split}\label{eq:DDW}
\end{equation}
\nid where $C$ is a constant to be specified. Additionally, for the sake of simplicity we considered that the bath couples to each degree of freedom independently, and the coupling function is given by 
\begin{equation}
\begin{split}\label{eq:coupling_f2}
f(q_1,q_2) = (f(q_1),f(q_2))
\end{split}
\end{equation}
\nid where $f(q_i)$ is given by Eq.~\ref{eq:coupling_f}. 

\subsection{Fcc metals}

We also perform first-principles atomistic simulations to showcase the methodology developed in this paper. More specifically, we focus on hydrogen and deuterium hopping reactions within interstitial sites in different bulk fcc metals: Pd, Pt, Cu, and Ag. The bulk systems were modelled by 2 $\times$ 2 $\times$ 2 cubic supercells  containing one hydrogen or deuterium atom and 32 metal atoms.
Energies and forces were computed employing density-functional theory (DFT) using the FHI-aims \cite{FHI-AIMS} code and the Perdew, Burke, and Ernzerhof (PBE) exchange-correlation functional \cite{PBE}. 
Geometries were relaxed using the standard \textit{light} settings (in the case of Pd, increasing the radial multiplier to 2) from FHI-aims until all forces were below 10$^{-3}$ eV/\AA. Minimum-energy pathways (MEP) were obtained with the string method\cite{Weinan_JCP_2007} combined with
the climbing image technique\cite{Graeme_JCP_2000} as implemented in the aimsChain package provided with the FHI-aims code. The BFGS algorithm was used as the optimization procedure  and the residual forces converged below 10$^{-3}$ eV/\AA.
Unless specified otherwise, a 6 $\times$ 6 $\times$ 6 k-point sampling was used.  This setup ensures that errors in relative energies are below 1~meV/atom. We  obtained lattice constants of
3.95~\AA, 3.97~\AA, 3.63~\AA, and 4.16~\AA,   for
Pd, Pt, Cu and Ag, respectively, 
in good agreement with Ref.~\cite{Haas_PRB_2009}.
See section III A of the SI for more details regarding convergence tests.

As explained in part I of this paper, the electronic friction tensor was computed as
\begin{equation}
\begin{split}
\tilde{\eta}_{ij}(\textbf{q},\lambda)&=  
\hbar\sum_{\nu,\nu'}
\braket{\psi_{\nu} |\partial_i \psi_{\nu'}} 
\braket{\psi_{\nu'} | \partial_j \psi_{\nu}}
(f(\epsilon_\nu) - f(\epsilon_{\nu'}))\\
&\times
\frac{\lambda \Omega_{\nu\nu'} }{\lambda^2 + \Omega_{\nu\nu'}^2},
\label{eq:eta-lorentz}
\end{split}
\end{equation}
where $f(\epsilon)$ is the state occupation given by the Fermi-Dirac distribution,
$\psi_\nu$ and $\epsilon_\nu$ are the KS electronic orbitals  and orbital energies of the $\nu$-th level, $i$ and $j$ label the nuclear degrees of freedom, $\partial_i = \partial/\partial q_i$, and $\Omega_{\nu\nu'}=(\epsilon_\nu -\epsilon_{\nu'})/\hbar$. The calculation of the non-adiabatic coupling elements was obtained through a finite-difference approach, as currently implemented in the FHI-aims code~\cite{Maurer_2016_PRB}. We stress that Eq. \ref{eq:eta-lorentz} is  different from the expression 
commonly used in the literature~\cite{Dou_PRL_2017,Maurer_2016_PRB,Head_Gordon_1995,Luntz_JCP_2005,Monturer_PRB_2010}.
We used a step length of 0.001~\AA~ for the finite-difference evaluation,  
and a 16 $
\times$ 16 $\times$ 16 k-point sampling for the friction tensor. We only calculated the tensor components related to the hydrogen or deuterium atoms.  More details regarding convergence tests can be found in section III. B of the SI.

\section{Numerical results on model potentials}

\subsection{Position-independent friction}\label{sec:res:DW}

\begin{figure}[h]
\centering
       \includegraphics[width=1.0\columnwidth]{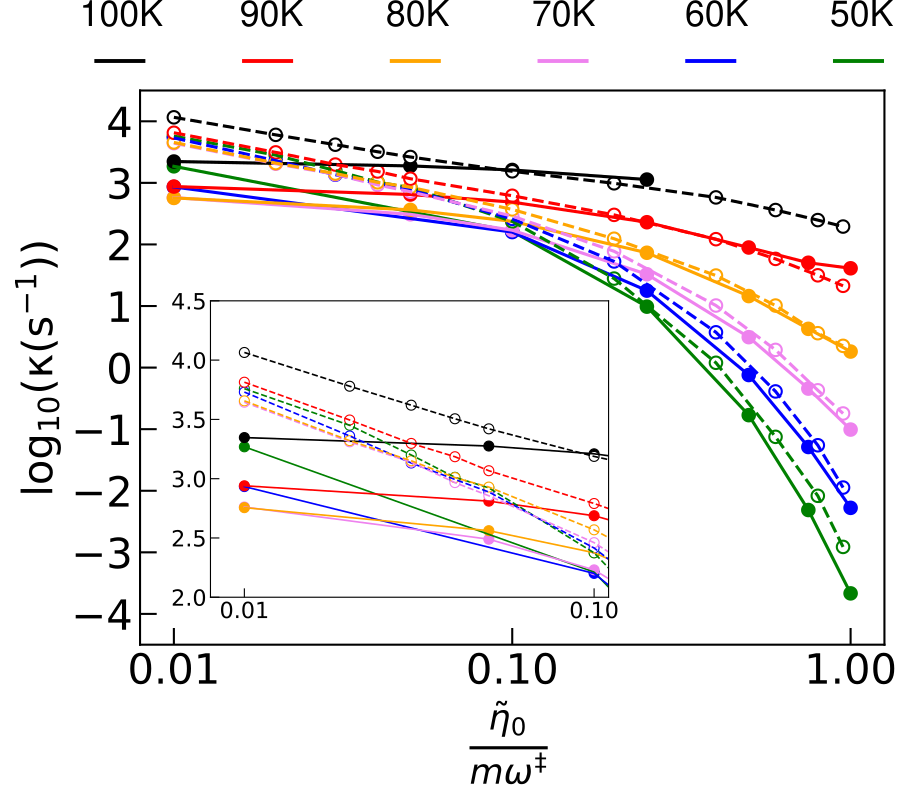}
    \caption{ Reaction rate constants for the DW potential ($V_0=258$ meV, $\omega_c=500$ cm$^{-1}$ and $q_0=0$) for temperatures between 50 K and 100 K for different friction values computed with RPI-EF (solid lines with filled circles) and ML-MCTDH (dashed lines with empty circles). }
    \label{fig:SI-RPI-MCTDH}
\end{figure}

We start by analyzing the linear coupling case and benchmarking the  RPI-EF results against the ML-MCTDH results. In Fig.~\ref{fig:SI-RPI-MCTDH} the rate constants calculated for the DW potential (Eq.~\ref{eq:DW}) at different temperatures and friction values are shown. The RPI-EF results are in good agreement with the exact calculations for $\tilde{\eta}_0/m \omega^\ddag > 0.1$ at all the temperatures considered. We note that the lowest temperature, 50 K, represents less than half of the cross-over temperature evaluated without friction. Even at considerably higher friction values, where the RPI approach has been predicted 
to be inadequate~\cite{Richardson_JCP_2009}, the agreement is quite remarkable, showing  
that real-time dynamical effects, such as recrossing, play a minor role in these cases.
At lower friction values, specially below $\tilde{\eta}_0/m \omega^\ddag = 0.05$,  the agreement deteriorates. The poor performance of the RPI-EF method in the weak coupling regime is not surprising since RPI predicts a finite value for the rate even at $\tilde{\eta}_0=0$, where the dynamics in such a 1D model would be described by a Rabi oscillation and, strictly speaking, a rate process cannot be defined. Indeed, RPI-EF reaches a plateau at $\tilde{\eta}_0/m \omega^\ddag \approx 0.05$ which might be interpreted as \enquote{an intrinsic} dissipation inherent to that theory, as a consequence of considering  trajectories that bounce only once in the evaluation of the imaginary time  kernel \cite{Jor_review_2018}. When $\eta$ goes to zero, the exact results approach the limit of coherent tunneling dynamics, yielding ML-MCTDH rates that are larger than the ones calculated with RPI.
In passing, we note that both methods present a minimum of the rate at around 70 K in the weak-friction regime (see inset in Fig. \ref{fig:SI-RPI-MCTDH}), which differs from the low-temperature power law observed in metastable systems \cite{Grabert_PRL_1984,Grabert_PRB_1987}.  A deeper study of this subtle but interesting quantum effect is beyond the scope of the current work and will be the subject of future research.

\begin{figure}[htb]
\centering
       \includegraphics[width=0.80\columnwidth]{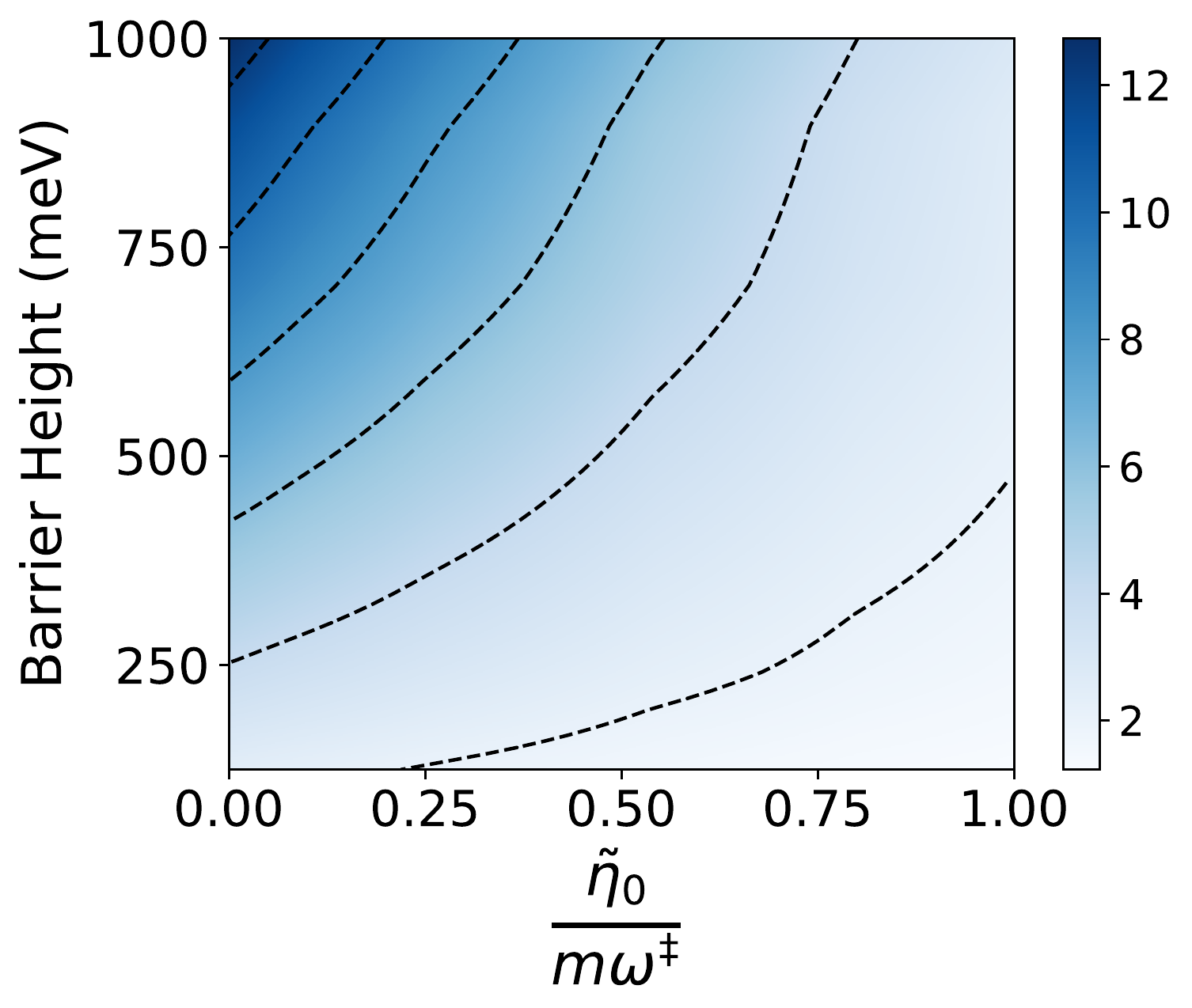}
    \caption{ Tunneling enhancement factors $\log_{10}[\kappa(\beta,\tilde{\eta}_0)]$ as a function of energy barrier height and friction strength, on the DW potential with $q_0=0.0$ \AA~ (symmetric reaction profile) at $T=0.7T^\circ_c=80$~K. The colour-scale  is logarithmic and contour lines are drawn for isosurfaces spaced by 2 logarithmic units.} 
    \label{fig:DW-sym}
\end{figure}

\begin{figure}[htb]
\centering
       \includegraphics[width=0.80\columnwidth]{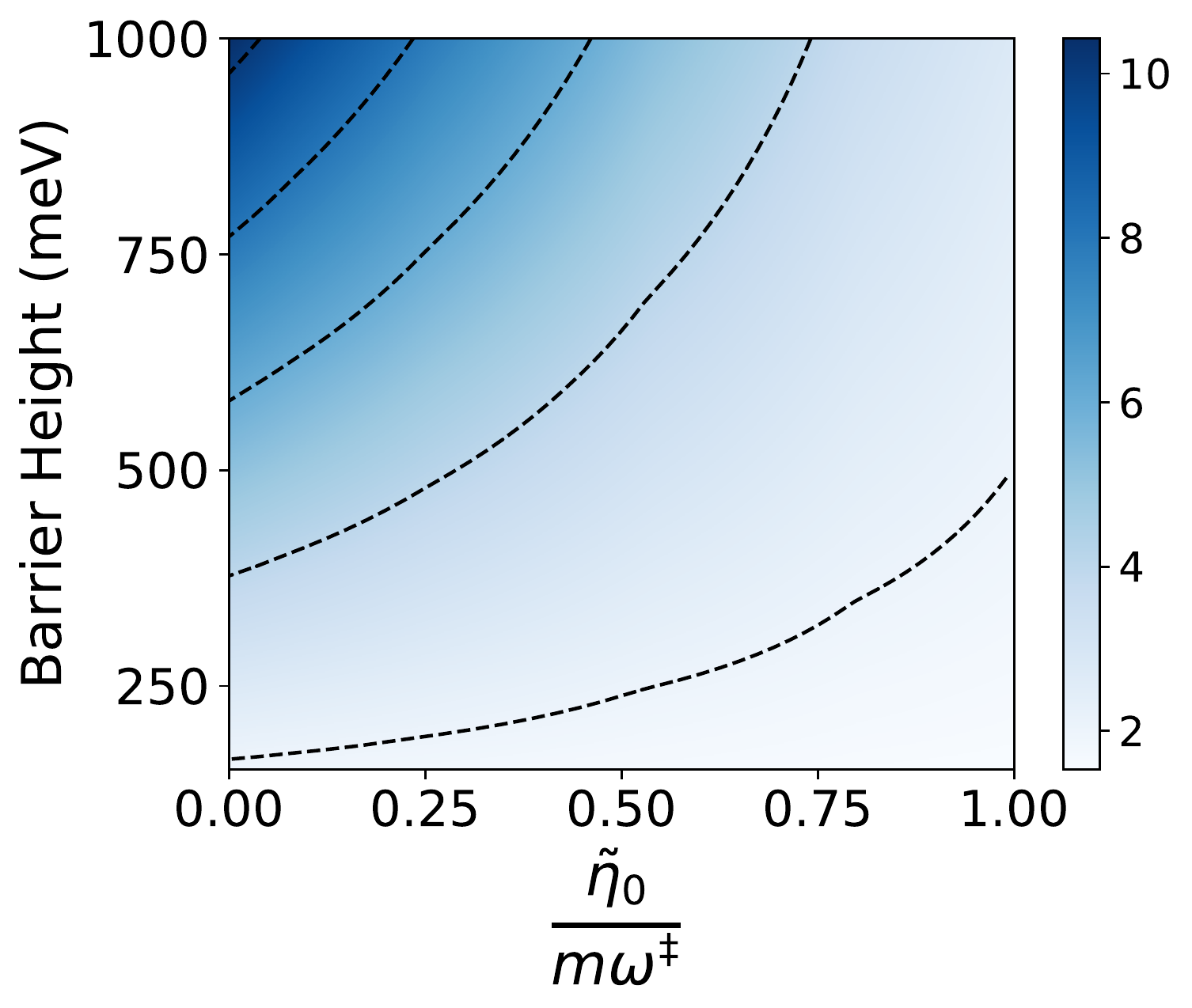}
    \caption{ Same as Fig. \ref{fig:DW-sym}  with $q_0=0.08$\AA~ (asymmetric reaction profile). The exoergic reaction is considered. } 
    \label{fig:DW-asym}
\end{figure}

One way to evaluate the relative importance of tunneling to the total rate is to analyze the tunneling enhancement factor, $\kappa^\text{tun}$, defined as 
\begin{equation}
\begin{split}\label{eq:kappa}
\kappa^\text{tun}(\beta,\tilde{\eta}) =  \frac{k_\text{inst}(\beta,\tilde{\eta})}{k_\text{TST}(\beta,\tilde{\eta})} 
\end{split}
\end{equation}
\nid with $k_\text{TST}(\beta,\eta)$ being the TST rate\cite{Beyer_PCL_2016}.
Figs. \ref{fig:DW-sym} and \ref{fig:DW-asym} show the calculated tunneling enhancement factors  for different barrier heights and friction values for symmetric and asymmetric reactions, respectively.
 For the range of parameters considered here, it can be observed that the 
 the tunneling enhancement factor calculated with and without friction can differ up to almost ten orders of magnitude, that it increases with the increase of the barrier height, and that it decreases with increasing friction strength.
 For barriers larger than 500 meV, the isosurfaces are approximately straight lines with slope one, meaning that barrier heights and friction strengths have a comparable but opposite effect on tunneling.

\begin{figure}[htb]
    \noindent
    \centering
    \includegraphics[width=\columnwidth]{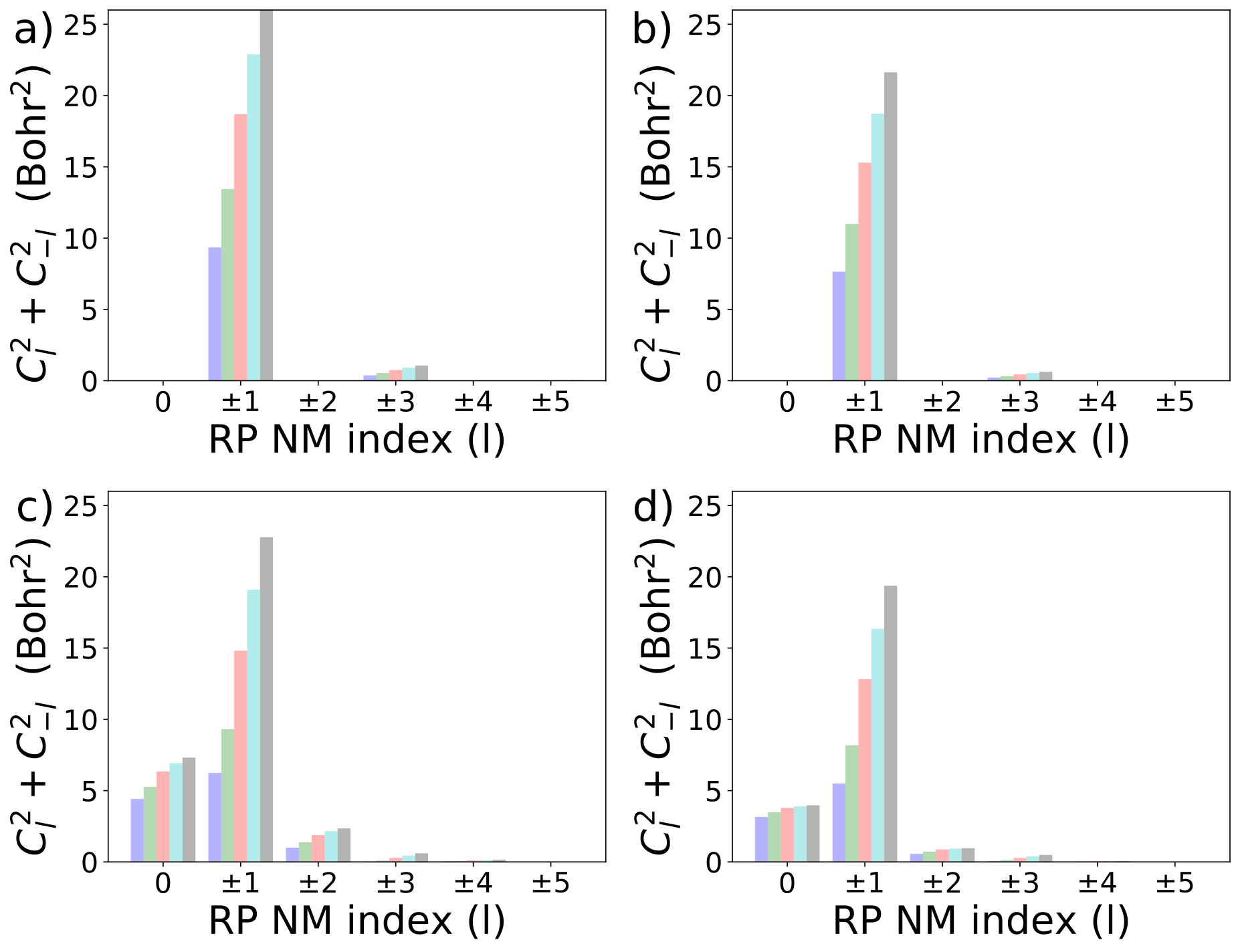}
    \caption{Decomposition of instanton geometry into the free RP normal mode  basis at a representative temperature of $T=0.70T^\circ_c=80$ K for (a) symmetric DW with $\tilde{\eta}_0/m\omega^\ddag=0.00$,
    (b) symmetric DW with $\tilde{\eta}_0/m\omega^\ddag=0.50$,
    (c) asymmetric DW ($q_0=0.08$\AA) with $\tilde{\eta}_0/m\omega^\ddag=0.00$, and
    (d) asymmetric DW ($q_0=0.08$\AA) with $\tilde{\eta}_0/m\omega^\ddag=0.50$.
    Five different barrier heights were considered: 125 meV (purple), 258 meV (green), 500 meV (red), 750 meV (light blue), and 1000 meV (gray).
    Coefficients are ordered and grouped by their corresponding RP normal mode (NM) index (l). }
    \label{fig:normal_mode_proj}
\end{figure}

We now proceed to discuss why the friction strength and the barrier height impact the tunneling contribution to the rate. 
The friction value determines the system-bath coupling strength. As evidenced by $\kappa^\text{tun} \sim 1$ at higher friction values, the stronger the coupling, the more classical the system behaves. However, the reason why the impact of the friction becomes more relevant at higher barrier heights  
is less straightforward to understand and requires the analysis of the instanton pathways. In Fig. \ref{fig:normal_mode_proj}, we show the decomposition of the instanton geometry into the free RP normal mode basis. We first consider the case without coupling to the bath. For symmetric barriers, Fig. \ref{fig:normal_mode_proj}a, the instanton pathway expands only along the odd RP normal modes due to the symmetry of the underlying potential, and the first two degenerate RP normal modes ($l=\pm1$) contribute with more than 99$\%$ to the path. For asymmetric barriers, Fig. \ref{fig:normal_mode_proj}c, even though all normal modes are in principle allowed by symmetry, the first two degenerate RP normal modes exhibit the highest contribution, with the centroid mode ($l=0$) presenting a non-negligible contribution as well. For both barrier shapes, the population of the $l=\pm1$ modes increases with the barrier height, simply because the pathway from reactants to product becomes longer. The same trends are observed for calculations with intermediate system-bath coupling, Fig. \ref{fig:normal_mode_proj}c and \ref{fig:normal_mode_proj}d, where the only difference is an overall smaller population of the RP normal modes due to the shorter instanton pathway. 
 Thus, higher barriers correlate with a relatively larger impact of the friction on the rate. This is due to an increase in the RP normal mode population, which leads to an increase of the last term in Eq. \ref{eq:MF-SI}.

Grote-Hynes (GH) theory~\cite{Grote_JCP_1980,Grote_JCP_1981,Pollak_JCP_1986} defines a relationship between reaction rates obtained with a finite friction strength and those obtained with vanishing friction strength in the classical limit. In part I  of this article, we showed an extension of GH theory to the deep tunneling regime for the case of position-independent friction. 
Briefly, we proposed that tunneling rates with finite friction strength -- RPI-EF rates -- at a given temperature, can be related to tunneling rates without friction -- RPI rates -- performed at a scaled temperature. The scaling relation for the temperatures is given by 
\begin{equation}
\begin{split}\label{eq:Scaling}
 T_b/T_c^{\circ} = \sqrt{
 \bigg(\frac{ \tilde{\eta}(\omega_l^b)}{2m{\omega^\ddag}}\bigg)^2 \frac{1}{l^2}+ \left(\frac{T_a}{T_c^{\circ}}\right)^2
 } - 
 \frac{\tilde{\eta}(\omega_l^b)}{2m{\omega^\ddag}}\frac{1}{l}
\end{split}
\end{equation}
\nid where $T_c^{\circ}$ is the cross-over temperature without friction, $T_b$ the target temperature at which the RPI-EF result is desired, $T_a$ the temperature at which the RPI calculation must be performed, and $\omega_l^b$ the $l$ free RP normal mode frequency at $T_b$.  Since $\omega^b_l$ depends on $T_b$, this equation has to be solved self-consistently. Even though Eq.~\ref{eq:Scaling} must be fulfilled for all $l$, from Fig.~\ref{fig:normal_mode_proj} one can expect that considering only $\omega_{l=\pm1}$ should be a good assumption.

\begin{figure}[htb]
    \noindent
    \centering
    \includegraphics[width=0.9\columnwidth]{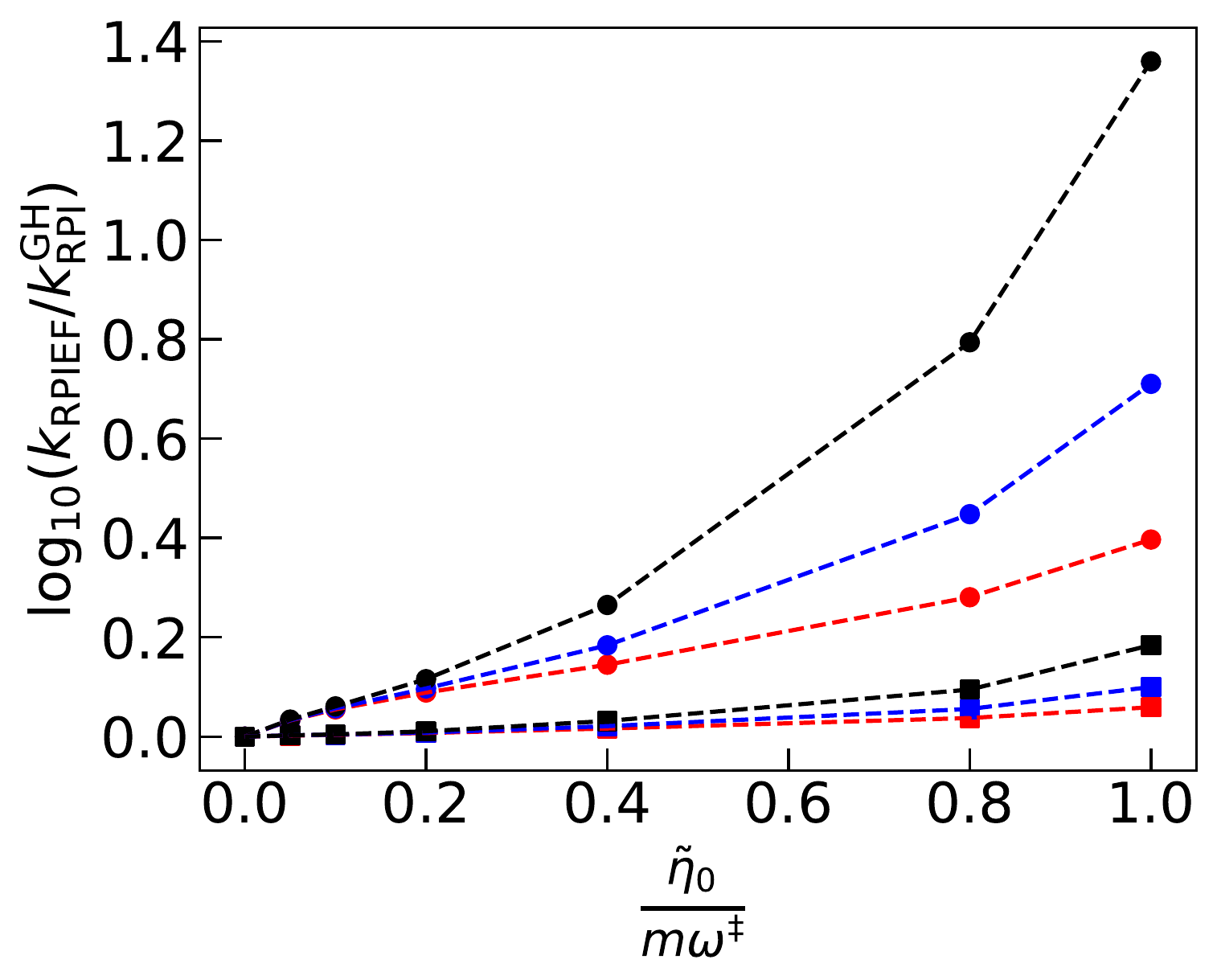}
    \caption{Error on the RPI rate values obtained by the scaling relation shown in Eq.~\ref{eq:Scaling} ($k_{\text{RPI}}^{\text{GH}}$), compared to RPI-EF rates ($k_{\text{RPIEF}}$). The error is reported as the logarithm of the ratio between these rates in the symmetric DW model. Temperatures of  0.70$T^\circ_c$ (squares) and
    0.55$T^\circ_c$ (circles), and reaction barriers  of 258 meV (red), 500 meV (blue), and 1000 meV (black) are shown. An analogous plot for an asymmetric barrier is presented in the SI. }
    \label{fig:rates_estimation}
\end{figure}

In Fig. \ref{fig:rates_estimation}, the error obtained by computing the rate, using only $\omega_{l=\pm1}$ in Eq.~\ref{eq:Scaling}, for different temperatures and coupling strengths in the symmetric DW potential is presented. The estimated RPI rates at the scaled temperatures are within one order of magnitude from the full RPI-EF rates for all friction strengths, but they are in better agreement for $\tilde{\eta}_0/m\omega^{\ddag} < 0.5$. Similar accuracy is observed for an asymmetric DW model with this approximation, even though the $l=0$ mode is appreciably activated (see section IV in the SI).

\subsection{Position-Dependent Friction}\label{sec:res:DDW}

\begin{figure}[htb]
\centering
       \includegraphics[width=0.80\columnwidth]{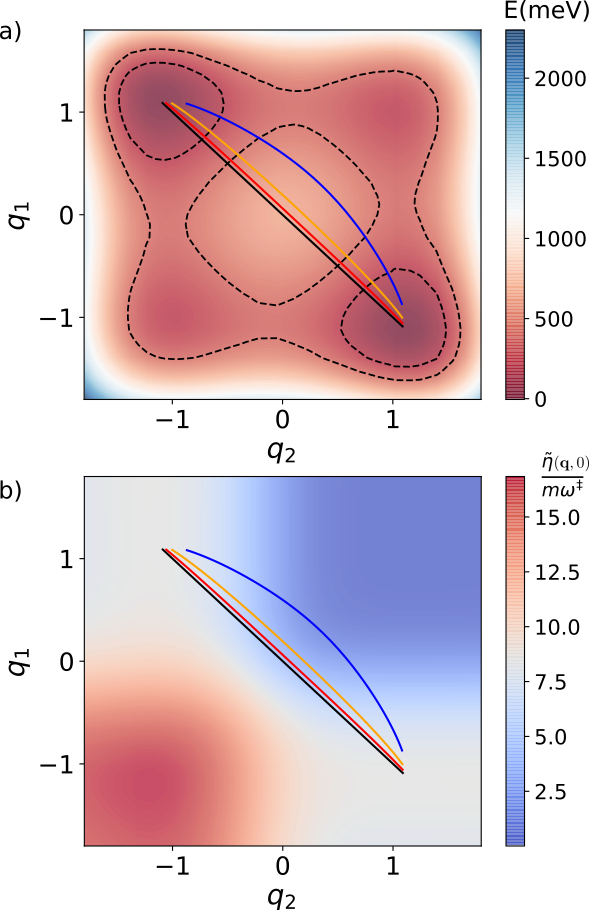}
   \caption{Instanton pathways obtained at 40 K using the DDW model with $V_0 = 258$ meV, $q_0 = 0$ \AA, $C=97.1$ meV/\AA$^2$ ( 0.001 a.u.), $\epsilon_1$ = 0, $\epsilon_2 = -0.8$, and $\Delta = 1.0$ for $\tilde{\eta}_0/m\omega^\ddag$= 0, 0.10, 0.25, and 0.50 represented by full black, red, orange, and blue lines, respectively. The pathways are shown on top of a) a heat map representing the underlying potential energy surface b) a heat map that helps visualize the position-dependence of the friction tensor. The map is computed as the sum of the diagonal elements of $\tilde{\eta}$ for $\tilde{\eta}_0/m\omega^\ddag=0.50$ (see Eq. \ref{eq:eta_laplace} and \ref{eq:coupling_f2} in the main text).}
    \label{fig:DDW}
\end{figure}

We now consider the case where the coupling between the system and the bath depends on the  position of the system coordinate, i.e. a position-dependent friction. Fig.~\ref{fig:DDW} shows instanton pathways obtained for different system-bath coupling strengths
on a DDW potential, as described by Eq.~\ref{eq:DDW}, at a temperature considerably lower than $T_c^\circ$ in this model.  The parameters of the model were chosen such that the PES presents two global minima, two local minima, four first order saddle points that connect each global minimum with the closest local minimum, and one second order saddle point at $(q_1,q_2)=(0,0)$, as shown in Fig. \ref{fig:DDW}a. The optimal tunneling pathway in the absence of dissipation is represented by the black curve in Fig.~\ref{fig:DDW} and it is a linear trajectory that connects the two global minima by crossing the second order saddle point. For $\tilde{\eta} >0$, the anisotropy of the friction (see Fig. \ref{fig:DDW}b) results in a modification of the instanton pathway, which bends towards regions of lower friction values. The magnitude of the bending of this path increases as the strength of the friction becomes larger. This shows that the optimal dissipative tunneling pathway is a compromise between the path with the shortest length, the path with the lowest potential energy, and the path with lowest friction.
Indeed, for  $\tilde{\eta}_0/m\omega^\ddag >0.5$, the dissipation is so strong close to the second-order saddle point that no instanton pathway that connects directly the two global minima can be found.


\section{Hopping of Hydrogen and Deuterium in Bulk Metals}\label{sec:res:ab_initio}

\subsection{Minimum energy paths, barrier heights and friction strengths}\label{sec:res:friction-static}

Having characterized the performance of RPI-EF in model potentials, we now address the interplay of NAEs and NQEs on the hopping reaction of H within bulk metals, which we calculate from first-principles electronic-structure simulations.  

We first analyze the MEP of the reactions in Pd, Pt, Cu, Ag and Al. We note that we focus on the hopping reaction between neighboring  octahedral$\rightarrow$tetrahedral interstitial sites. Even though in perfect solids these reactions  determine the diffusion rate, in real materials other mechanisms might become the rate determining step of the diffusion process~\cite{DISTEFANO_JCP_2015,PEDERSEN_Acta_2009}.  
In Table \ref{tab:Metals}, we report the reaction energy, reaction barrier, and electronic-friction values along the MEP for the reactions considered in this work. 
As shown in column 3, the energy barriers are in the 100-300 meV range, in accordance with previous studies \cite{Ferrin_SurfSci_2012}.

\begin{table}[h]
    \centering
    \begin{tabular}{c|c|c|c|c|c|c|c}
         System    &  $E_{\text{T-O}}$ (meV) & $E_{\text{TS-O}}$ (meV) & 
         $\tilde{\eta}$ (ps$^{-1}$) & $T_c^{\circ}$ (K)& $\omega^\ddag$ (cm$^{-1}$)\\
         \hline
H$@$Pd   &   43        & 148   &    0.7 - 2.7   & 115   & 501  \\
H$@$Pt   &   -35       &  44   &    0.8 - 2.8   & 96    & 420  \\
H$@$Cu   &   188       &  300  &    0.7 - 1.1   & 140   & 612  \\
H$@$Ag   &   52       &  160  &     0.7 - 1.0   & 116   & 504  \\
H$@$Al   &   -71      &  88   &     1.8 - 3.1   & 84    & 365  \\
    \end{tabular}
    \caption{ Reaction energy, $E_{\text{T-O}} = E_{\text{T}}-E_{\text{O}}$,     and energy barrier heights, $E_{\text{TS-O}} = E_{\text{TS}}-E_{\text{O}}$,
    for the different fcc metals considered in this work. $E_{\text{T}}$, $E_{\text{O}}$, and $E_{\text{TS}}$ refer to the potential energy corresponding to structures where the H atom is located at the tetrahedal (T), octahedral (O) and transition state (TS) sites, respectively. Values are reported without ZPE corrections. 
    Minimum and maximum values adopted by the electronic friction, $\tilde{\eta}$, along the MEP are presented in column four. Values are evaluated at 54 meV (which corresponds to the first non-zero ring-polymer normal mode frequency at 100 K) and considering a projection of $\tilde{\eta}$ on the reaction coordinate. Columns 5 and 6 show the crossover temperature, $T_c^\circ$, and imaginary frequency at the TS, $\omega^\ddag$, respectively. }
    \label{tab:Metals}
\end{table}

\begin{figure}
    \centering
    \includegraphics[width=\columnwidth]{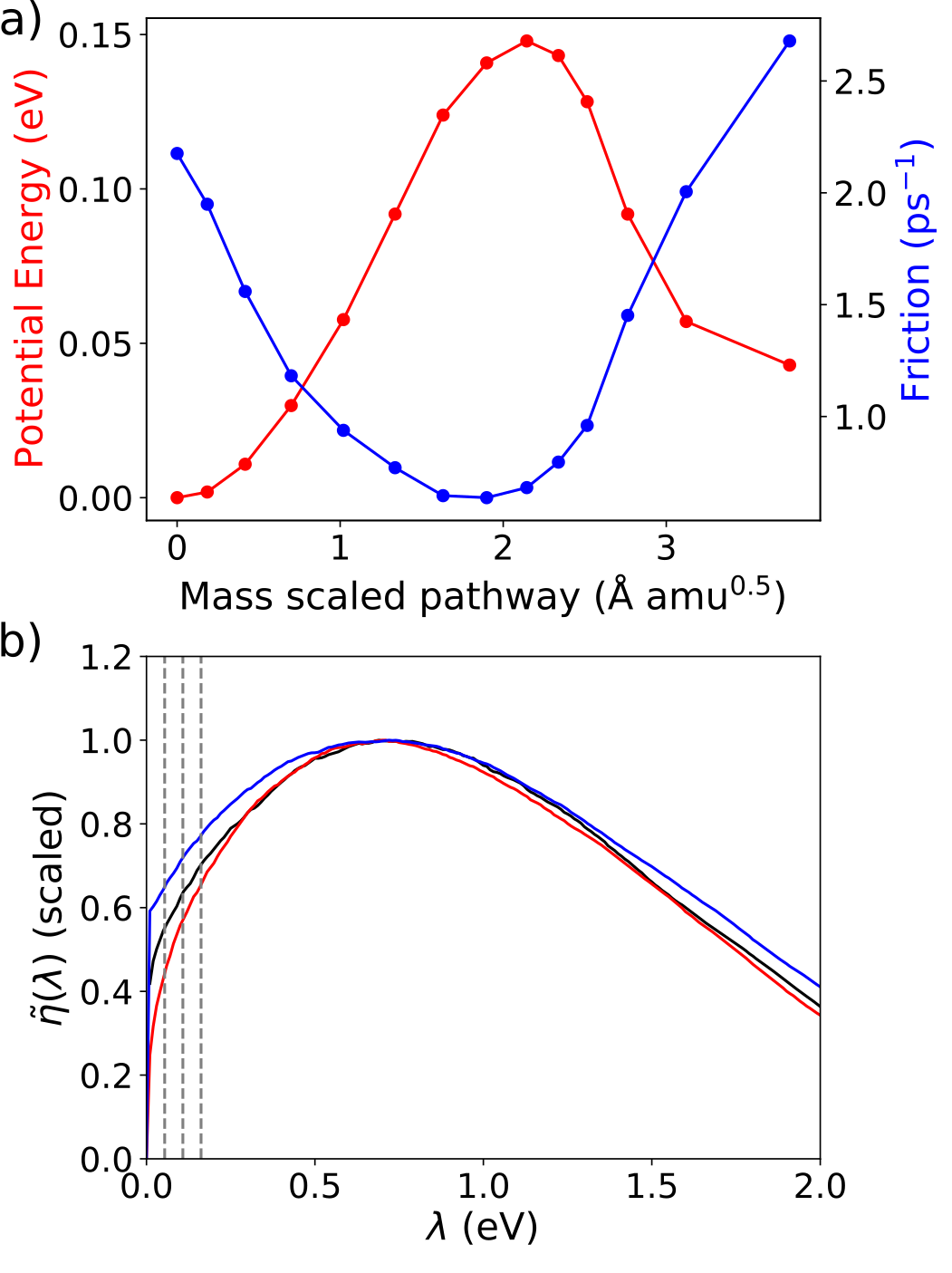}
    \caption{
    a) Minimum energy pathway (MEP) and friction along the reaction coordinate for the H hopping reaction in Pd. The  energy is set to zero at the reactant geometry.  
    b) Frequency dependence of the friction tensor  projected on the reaction coordinate at the reactant (black), transition state (red), and product (blue) states.    To ease visual comparison, all curves in panel b have been scaled to adopt the value of 1 at the highest friction value. The first three non-zero ring-polymer normal modes frequencies at 100K are  depicted as vertical dashed gray lines.
    }
    \label{fig:Pd32H}
\end{figure}

We continue by analyzing the electronic friction tensor $\tilde{\eta}_{il}(\bm{q}, \lambda)$ on the hydrogen atom along the MEPs. 
In Fig. \ref{fig:Pd32H}a, we present the electronic friction values evaluated at the first non-zero ring-polymer normal mode frequency at 100~K, projected onto the direction parallel to the reaction coordinate for the case of Pd. The friction values vary up to almost an order of magnitude, indicating the necessity of having a rate theory that takes into account such position dependence. 
The electronic friction along the MEP for the other metals shows a strong position dependence as well (see section III B of the SI).
As shown in column 4 of Table~\ref{tab:Metals}, the values vary from 0.7~ps$^{-1}$ to 3.1~ps$^{-1}$. The magnitude of the friction coefficients is large enough to impact vibrational lifetimes \cite{Maurer_2016_PRB}, adsorption mechanisms \cite{Bunermann_Sci_2015}, and scattering experiments \cite{Kandratsenka_PNAS_2018,Bunermann_Sci_2015}.  However, the dimensionless coefficient $\tilde{\eta}/m\omega^\ddag$ yields at most a value of 0.05 (for Al), which, given the relatively low barrier heights, would result in a reduction of the tunneling rates by less than a factor of 5, according to our study on model potentials presented in the previous sections (see Figs.~\ref{fig:DW-sym} and \ref{fig:DW-asym}).  


 A closer look at the expression used to compute the electronic friction tensor, Eq. \ref{eq:eta-lorentz}, allows us to rationalize the reasons behind such small coefficients \cite{Forsblom_JCP_2007}. Large friction values will arise in the case of a high DOS close to the Fermi level and due to the presence of hydrogen states close to the Fermi level. The former contributes to Eq.~\ref{eq:eta-lorentz} via the Fermi-Dirac factors, the latter contributes via the strength of the nonadiabatic coupling. 
 While the former depends mainly on the metal at hand, the latter is affected by both the impurity and the metal\cite{Lecroart_JCP_2021}. We analyzed the atomic projected DOS (see section III C in the SI) and confirmed that Pd and Pt are, at the same time, the systems that present the largest electronic friction values and the highest DOS at the Fermi level among the transition metals. Surprisingly, Al presents slightly larger friction values, without a high DOS at the Fermi energy. This might suggest that the nonadiabatic couplings are 
 comparatively large in this case. However, in all cases, the hydrogen atom neither creates new states nor affects the DOS appreciably in the vicinity of the Fermi level, which ultimately leads to rather small friction coefficients along the MEP for these systems. 
  
Up to this point, we have considered values of the friction tensor at a single frequency.
However, the calculation of the tunneling rates require the evaluation of the friction tensor at all the RP normal modes frequencies (Eq.~\ref{eq:eff-SD}), and more importantly in the derivation of RPI-EF with a position-dependent friction tensor, we have assumed that a separable coupling \textit{ansatz} is valid. This \textit{ansatz} is equivalent to assuming that the frequency dependence of the friction tensor remains the same at all (relevant) positions.
As an illustrative example, we present in Fig. \ref{fig:Pd32H}b the frequency dependence of the friction tensor at the stationary points of the MEP (reactant, transition and product state) for the hydrogen hopping reaction in Pd. 
The  frequency dependence shows a non-monotonic profile with a maximum around 0.6 eV 
and remains fairly similar along the MEP suggesting that the non-adiabatic couplings in bulk metals are, to a great extent, well described by a ``separable coupling".
This observation is equally valid for the other  metals (see SI section III B). 
A different type of coupling might be observed in scattering reactions, where atoms or molecules transition from vacuum to electron-rich  environments\cite{Bunermann_Sci_2015,Maurer_PRL_2017}.
As discussed in section \ref{sec:res:DW}, only the first few ring-polymer normal modes are appreciably activated. For this reason, only a relatively small fraction of spectral density contributes to the rates (see Fig. \ref{fig:Pd32H}b). 

\subsection{Tunneling rates: The case of Pd}\label{sec:res:Pd32HD}

We performed full-dimensional instanton calculations on Pd in order to gauge the predictive power of our studies on low-dimensional models and static estimators presented in the previous sections. 
We selected Pd because it presents high values of friction along the MEP and the diffusion of H in Pd has been well studied theoretically and experimentally before. In order to reduce the computational cost, we performed instanton calculations on a fcc cubic cell containing 4 Pd atoms and 1 H or D atom. These calculations were performed using a 12 $\times$ 12 $\times$ 12 k-point sampling. The relatively small size of the unit cell induces an effective increase of the barrier when compared to larger unit cells
and the new $T_c^{\circ}$ increases to 136 K (see section I in the SI). Since larger barriers magnify the effect of the friction on the rates, these calculations can be considered an upper-limit estimation of the impact of the friction on these rates. 

\begin{figure}[h]
\centering
       \includegraphics[width=1.0\columnwidth]
       {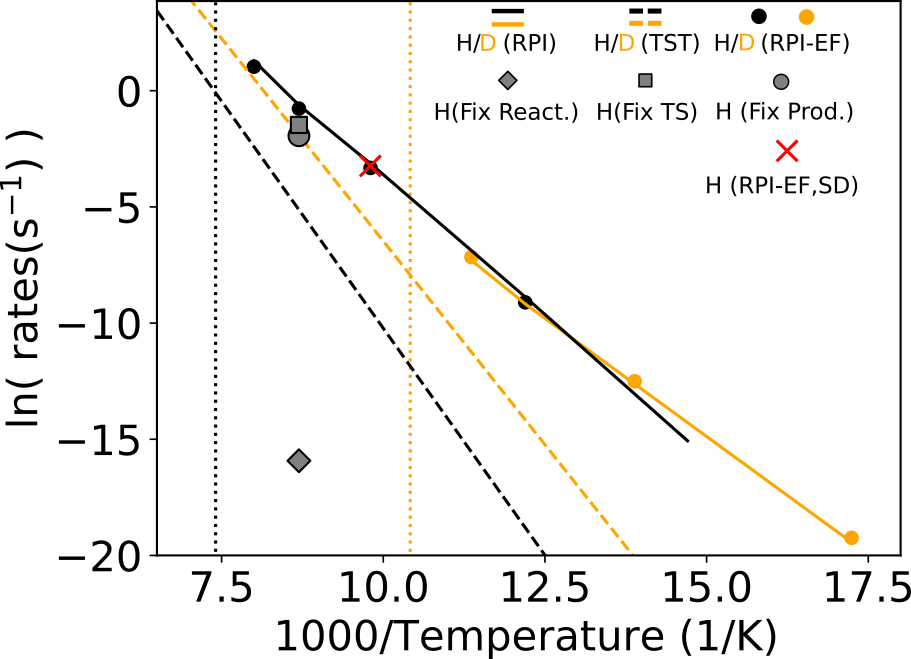}    \caption{Reaction rates for octahedral$\to$tetrahedral-site hopping reaction of H (black) and D (orange) in Pd, calculated by TST (dashed lines), and RPI rate theory (solid lines). RPI-EF rates with position-independent friction, fixed at a value of 4 ps$^{-1}$, are shown by black and orange circles for H and D, respectively, while  the RPI-EF rates with position-dependent friction is depicted as a red cross. RPI rate calculations with the Pd atoms fixed at their reactant, transition and product states are presented by diamond, square and circle gray symbols, respectively. $T_c^{\circ}$ for H and D are marked by vertical black and orange dotted lines, respectively.} 
    \label{fig:Pd-rates}
\end{figure}

In Fig.~\ref{fig:Pd-rates}, the reaction rates for the hopping reaction of H and D in Pd from the octahedral to the tetrahedral site using TST, RPI theory and RPI-EF theory are presented. At temperatures below $T_c^{\circ}$, the tunneling effects, evidenced by the difference between the RPI and TST rates, become increasingly important, enhancing the rate by several orders of magnitude. The comparison of the TST predictions for H and D indicates an inverse kinetic isotope effect (KIE). The inverse KIE can be traced back to the softening of the normal modes orthogonal to the reaction pathway at the transition-state geometry \cite{Meisner_Angew_2016}. Since this effect is mainly due to ZPE, it is also present at temperatures above $T_c$ and has been reported experimentally \cite{Volkl_book}. Below $T_c$, the emergence of tunneling creates a competition between  ZPE and tunneling effects as already reported by Shiga \textit{et al.}~\cite{Kimizuka_PRB_2019}. Moreover, the similarity of the reaction rates for both isotopologues around 80 K is in agreement with the results of Ref. \cite{Kimizuka_PRB_2019} and with experiments. However, the absolute values reported here are considerably smaller as a consequence of the rather small unit-cell employed in the calculations.

We now inspect the impact of the lattice on the reaction rates. For this purpose, we performed RPI calculations where the Pd atoms were fixed at the reactant, product and transition state positions at 115 K. The rate obtained when the Pd atoms are fixed at their reactant positions is 7 and 4 orders of magnitude slower than the RPI and TST estimates, respectively. This confirms the significant contribution of the lattice fluctuations to the reactive process \cite{Kimizuka_PRB_2018} and highlights  the multidimensional nature of the reaction. In contrast, the rate estimates obtained from the calculations with the Pd atoms fixed at their transition-state or  product  positions are within an order of magnitude of the RPI estimates. The former result is expected since we are analyzing the rates at only 20 K below $T_c^\circ$, such that the instanton pathway lies very close to the TS geometry. The latter result, instead, shows that the lattice relaxation between the transition state and the product is of comparatively lower relevance.

Finally, we consider the effect of the electronic friction on the rate. At 100 K, we performed a calculation with on-the-fly estimations of the friction tensor along the instanton pathway (also included in the instanton optimization). In order to gauge the importance of the spatial dependence of the friction, we also performed, at the same temperature, an instanton calculation using a constant value of 4 ps$^{-1}$  for the friction (slightly higher than the maximum value reported in Table \ref{tab:Metals}). The results were numerically indistinguishable within the accuracy of our calculations, which may seem an unexpected result, at first. We proceeded to perform calculations with a constant and spatially-independent friction value (which is less computationally demanding) at several other temperatures. In Figure \ref{fig:Pd-rates}, we show these results and, as predicted by our earlier assessment, the friction coefficients are not large enough to produce a significant effect on the rates. This explains why the spatial dependence also does not appreciably change the rate constants.

\section{ Conclusions }\label{sec:Conclusions}

We have benchmarked the RPI-EF method and showed its performance in model potentials and first-principles calculations. By performing numerically exact simulations in 1D model systems including a spatially-independent friction, we showed that RPI-EF yields accurate rates for all but very small friction coefficients at a much reduced computational cost.  A systematic analysis of 1D and 2D double-well potentials allowed us to determine the magnitude of the decrease of the tunneling rates caused by friction, as a function of the coupling strength and barrier height. 
We found that the suppression of tunneling is promoted by high coupling strengths and high energy barriers. We were also able to demonstrate that for a spatially-dependent friction tensor, the instanton pathway can be considerably deformed towards low-friction regions, when compared to the ``non-dissipative'' path (without friction).
 In comparison to previous similar approaches\cite{CALDEIRA_1983,Weiss_book}, the RPI-EF method is advantageous because it is  highly-efficient, more intuitive and mathematically simpler.

In the context of reactions involving atoms and molecules in metallic environments, RPI-EF allows the inclusion of NQEs and NAEs as described by an effective electronic friction.
While here we used an electronic friction formulation that disregards electronic correlation~\cite{Head_Gordon_1995,Dou_PRL_2017,Maurer_2016_PRB}, the RPI-EF approach is rather general and can be combined with other flavours of electronic friction that go beyond the independent quasi-particle picture.
As a consequence of the relatively low computational cost of both RPI and the employed \textit{ab initio} electronic friction formalism, RPI-EF allows the study of high-dimensional systems with on-the-fly \textit{ab initio} evaluation of the forces and the electronic friction tensor. 

In this work, we presented calculations of  hydrogen and deuterium hopping between nearest interstitial sites of selected fcc metals, employing density-functional theory calculations. By evaluating the impact that different factors have on the reactions rates of this reaction in bulk Pd, we established that nuclear tunneling and lattice relaxation play a larger role in determining the magnitude of the rate than electronic friction. The latter turned out to have a negligible impact on the reaction rates of these systems. This negative result, however, answers an important theoretical question regarding the interplay between NAEs (modeled by electronic friction) and NQEs\cite{Dou_JCP_2018}, and also validates the rate constants currently used in this context for multiscale modeling\cite{Huter_Metals_2018}.
Nonetheless, we anticipate that for impurities or adsorbates that present electronic levels in the vicinity of the Fermi energy of the metal, NAEs might play a more prominent role in connection with tunneling. We also expect to observe a larger effect for lighter particles such as Muons \cite{Muons2}, or for surface reactions with higher energy barriers \cite{GOMEZ_AppliedSurSci_2017}. 

\begin{acknowledgments}
Y.L., E.S.P. and M.R. acknowledge financing from the Max Planck Society and computer time from the Max Planck Computing and Data Facility (MPCDF).
Y.L and M.R. thank Jeremy Richardson, Aaron Kelly, and Stuart Althorpe for a critical reading of the manuscript.
C.L.B. acknowledges financial support through an EPSRC-funded PhD studentship. R.M. acknowledges Unimi for granting computer time at the CINECA HPC center. R.J.M. acknowledges financial support through a Leverhulme Trust Research Project Grant (RPG-2019-078) and the UKRI Future Leaders Fellowship programme (MR/S016023/1).

\end{acknowledgments}

\bibliographystyle{aipnum4-1}

\end{document}



\title{SUPPORTING INFORMATION:\\Dissipative Quantum Tunneling Rates through the Incorporation of First-Principles Electronic Friction in Instanton Rate Theory}

\author{Y. Litman}
\email{yairlitman@gmail.com}
\affiliation{MPI for the Structure and Dynamics of Matter, Luruper Chaussee 149, 22761 Hamburg, Germany}

\author{E. S. Pós }%
\affiliation{MPI for the Structure and Dynamics of Matter, Luruper Chaussee 149, 22761 Hamburg, Germany}

\author{C. L. Box}%
\affiliation{Department of Chemistry, University of Warwick, Coventry CV4 7AL, United Kingdom}

\author{R. Martinazzo}
\affiliation{Department of Chemistry, Università degli Studi di Milano, Via Golgi 19, 20133 Milano, Italy}

\author{R. J. Maurer}%
\affiliation{Department of Chemistry, University of Warwick, Coventry CV4 7AL, United Kingdom}

\author{M. Rossi}
\email{mariana.rossi@mpsd.mpg.de}
\affiliation{MPI for the Structure and Dynamics of Matter, Luruper Chaussee 149, 22761 Hamburg, Germany}

\date{\today}

\maketitle

\clearpage

\section{Convergence of  Instanton Calculations}

The convergence tests for the tunneling rates were performed using a 2x2x2 orthorhombic supercell containing four Pd atoms since it is expected to present similar convergence behavior as an 8x8x8 orthorhombic supercell containing 32 Pd atoms. In Table \ref{tab: Pd04H}  the tunneling   rates with varying number of beads is presented.  We analyze the
$PT/T_c$ ratio where $P$ is the number of beads, $T$ the temperature and $T_c$ the cross-over temperature as suggested in Ref. \cite{Litman_thesis}. We also report the $T_c$ for H and D in tables \ref{tab:TcH} and \ref{tab:TcD}. It can be seen that for all the temperatures considered, $PT/T_c$ ratios of 36 and 72 deliver converged rates within 
    10\% and 2\%, respectively. 

\begin{table}[h!]
    \centering
    \begin{tabular}{c|c|c|c|c|c|c|c|}
$T$(K) & $T/T_c$ &  $P$  & $PT/T_c$ & $k_\text{inst}$ (s$^{-1}$) \\
\hline
125 & 0.92   & 32  & 29.4 & 3.17      \\
125 & 0.92   & 96  & 88 &   3.66      \\
\hline 
115 & 0.85   & 32  & 27.2 & 6.44(-1)  \\
115 & 0.85   & 64  & 54.4 & 5.50(-1)  \\
115 & 0.85   & 96  & 82   & 5.38(-1)  \\
115 & 0.85   & 128 & 108  & 5.34(-1)  \\
115 & 0.85   & 256 & 216 & 5.39(-1)   \\
 \hline
102 & 0.75   & 48  & 36 & 4.74(-2)    \\
102 & 0.75   & 96  & 72 & 4.39(-2)    \\
102 & 0.75   & 128  & 96 & 4.33(-2)   \\
 \hline
82 & 0.60   & 64   & 38  & 1.55(-4)   \\
82 & 0.60   & 128  & 76  & 1.41(-4)   \\
82 & 0.60   & 160  & 96  & 1.40(-4)   \\
 \hline
68 & 0.50   & 72   & 36 & 3.04(-7)    \\
68 & 0.50   & 144  & 72 & 2.87(-7)    \\
68 & 0.50   & 192  & 96 & 2.85(-7)    \\
    \end{tabular} 
    \caption{ Tunneling Rates for \ce{Pd04H}.  } 
    \label{tab: Pd04H}
\end{table}

\begin{table}[h!]
    \centering
    \begin{tabular}{c|c|c|c|c|c|c|c|}
System & $T_c$ (K) & $\omega_\text{TS}$ (cm$^{-1}$)  \\
\hline 
\ce{Pd04H} & 136 & 594    \\
\ce{Pd16H} & 123 & 537    \\
\ce{Pd32H} &  114   & 502    \\
    \end{tabular} 
    \caption{ Cross-over temperature ($T_c$) and imaginary frequency at the transition state ($\omega_\text{TS}$).
    \ce{Pd32H}, \ce{Pd16H}, and \ce{Pd04H}, were computed with a k-grid mesh of
    6x6x6,
    6x6x6,
    and 12x12x12,
    respectively.
    }
    
    \label{tab:TcH}
\end{table}

\begin{table}[h!]
    \centering
    \begin{tabular}{c|c|c|c|c|c|c|c|}
System & $T_c$ (K) & $\omega_\text{TS}$ (cm$^{-1}$)  \\
\hline 
\ce{Pd04D} &     96     &   422   \\
\ce{Pd16D} &     87     &   381  \\
\ce{Pd32D} &     82     &   356  \\
    \end{tabular} 
    \caption{ Cross-over temperature ($T_c$) and imaginary frequency at the transition state ($\omega_\text{TS}$).
    \ce{Pd32D}, \ce{Pd16D} and \ce{Pd04D}, were computed with a k-grid mesh of
    6x6x6,
    6x6x6,
    and 12x12x12,
    respectively.
    }
    \label{tab:TcD}
\end{table}

\clearpage

\input{SI-MCTDH}

\section{Convergence Tests of DFT calculations}

\subsection{Energy Barriers}

\begin{figure}[h!]
    \noindent
    \centering
    \includegraphics[width=.8\columnwidth]{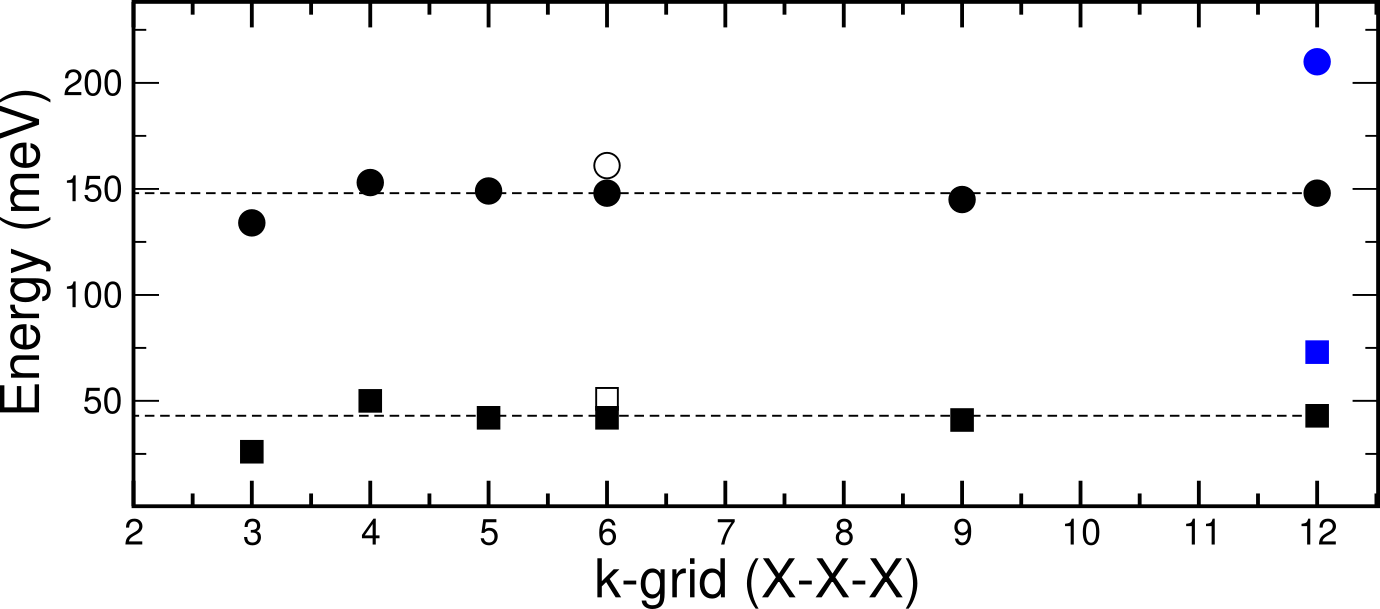}
    \caption{Reaction energy, $E_{\text{T}}-E_{\text{O}}$,     and energy barrier heights, $E_{\text{SP}}-E_{\text{O}}$,
    for  Pd are represented by squares and circles, respectively. $E_{\text{T}}$, $E_{\text{O}}$, and $E_{\text{SP}}$ refer to the potential energy corresponding to structures where the H atom is located at the tetrahedal (T), octahedral (O) and saddle-point (SP) sites, respectively. Values reported without  ZPE correction. Calculations were performed using supercells constructed from 2x2x2 (black) and 1x1x1 (blue) copies of the cubic unit cell. Dashed lines are shown as a guide to the eye. Standard \textit{light} with a modified radial multiplier=2 (filled symbols) and \textit{tight} (empty symbols) settings from FHI-aims were used.
    }     \label{fig:energy-kgrid}
\end{figure}

In Fig. \ref{fig:energy-kgrid} the convergence of the energy with respect to k-grid sampling for the hydrogen hopping reaction in Pd is shown. It can be observed that a 6x6x6 k-grid with \textit{light} settings delivers converged results for the reaction energy and energy barrier for the 2x2x2 supercell. The corresponding values obtained for the 1x1x1 supercell show a considerably larger reaction energy and barrier height.

\subsection{Friction Calculations}

\subsubsection{k-grid Convergence}

\begin{figure}[h!]
    \noindent
    \centering
    \includegraphics[width=1.0\columnwidth]{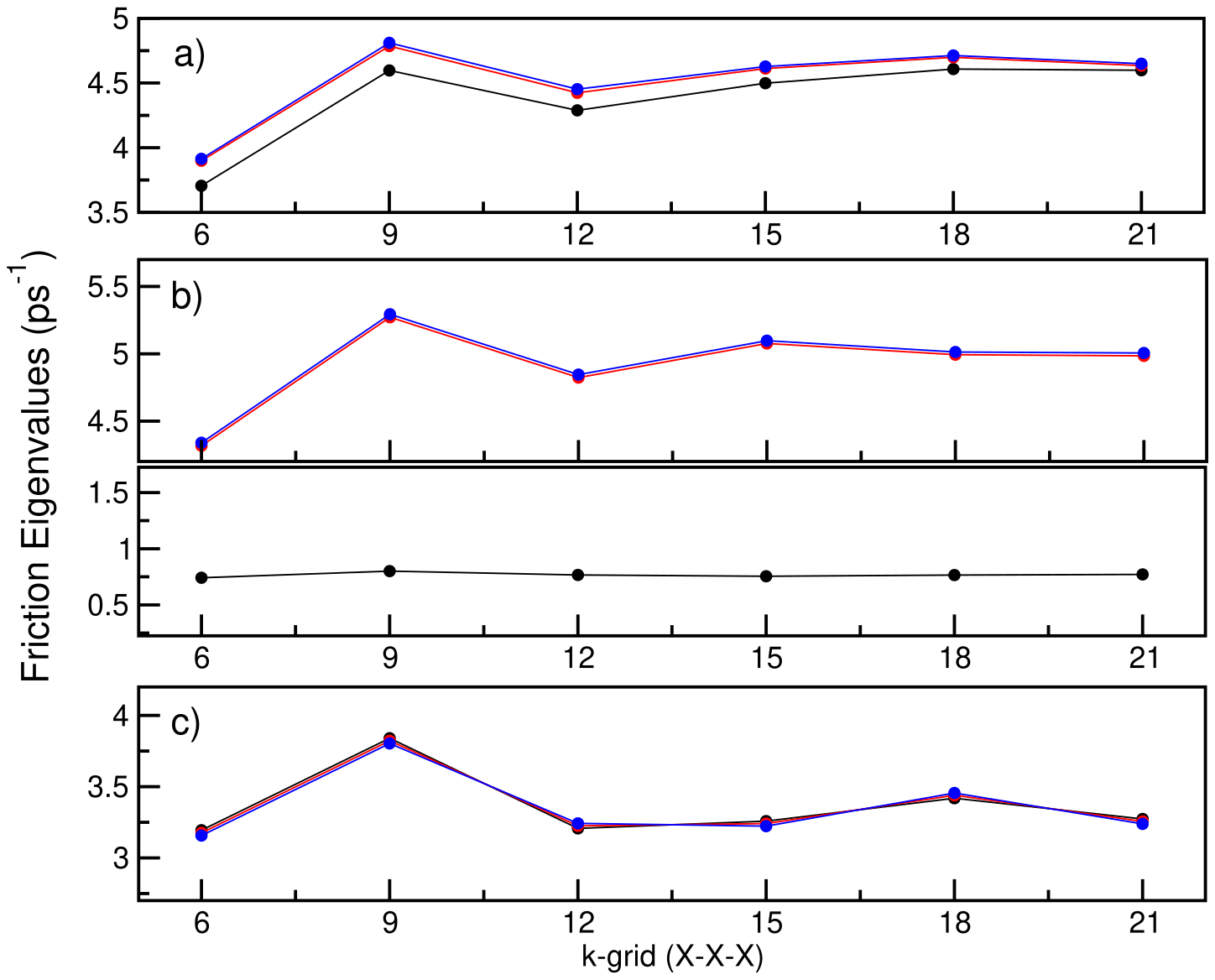}
    \caption{Convergence of eigenvalues of the friction tensor on the hydrogen atom for \ce{Pd32H} for  a) tetrahedral site, b) transition state between tetrahedral and octahedral sites, and c) octahedral site. Standard \textit{light} from FHI-aims were used with a modified radial multiplier=2 (filled symbols) were used. A broadening of 0.05 eV was used and the reported values are evaluated in the static limit. The different colours represent the three eigenvalues (one unique and two degenerate).}
    \label{fig:friction-kgrid}
\end{figure}

Fig. \ref{fig:friction-kgrid} shows the convergence of the friction tensor on the hydrogen on different positions with respect to k-grid sampling.  The friction eigenvalues converged within 20
\% and 3 \% with k-grid of 6x6x6 and 15x15x15, respectively. 

\subsubsection{Displacement Convergence}
\begin{figure}[h!]
    \noindent
    \centering
    \includegraphics[width=.9\columnwidth]{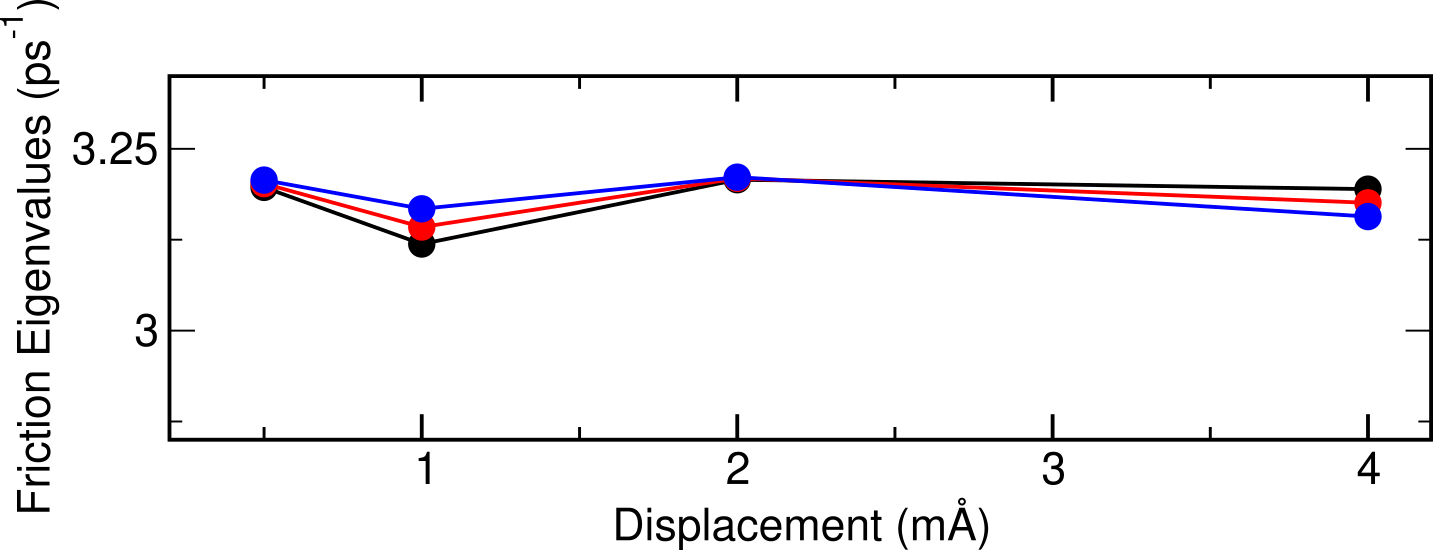}
    \caption{ Convergence of eigenvalues of the friction tensor on the hydrogen atom  with respect to displacement length for the hydrogen atom on the octahedral site (\ce{Pd32H}). A k-grid of 6x6x6 has been used and other settings were set equal to the k-grid convergence test.}     \label{fig:friction-displacement}
\end{figure}

Fig. \ref{fig:friction-displacement} shows the convergence of the friction tensor with respect to the displacement length used in the finite difference calculation. The friction eigenvalues vary by less than 5\% in the 0.5$\times$10$^{-3}$ to 4$\times$10$^{-3}$ \AA~range.

\clearpage

\subsubsection{Friction along the Minimum Energy Pathways}

\begin{figure}[h!]
    \centering
    \includegraphics[width=0.7\columnwidth]{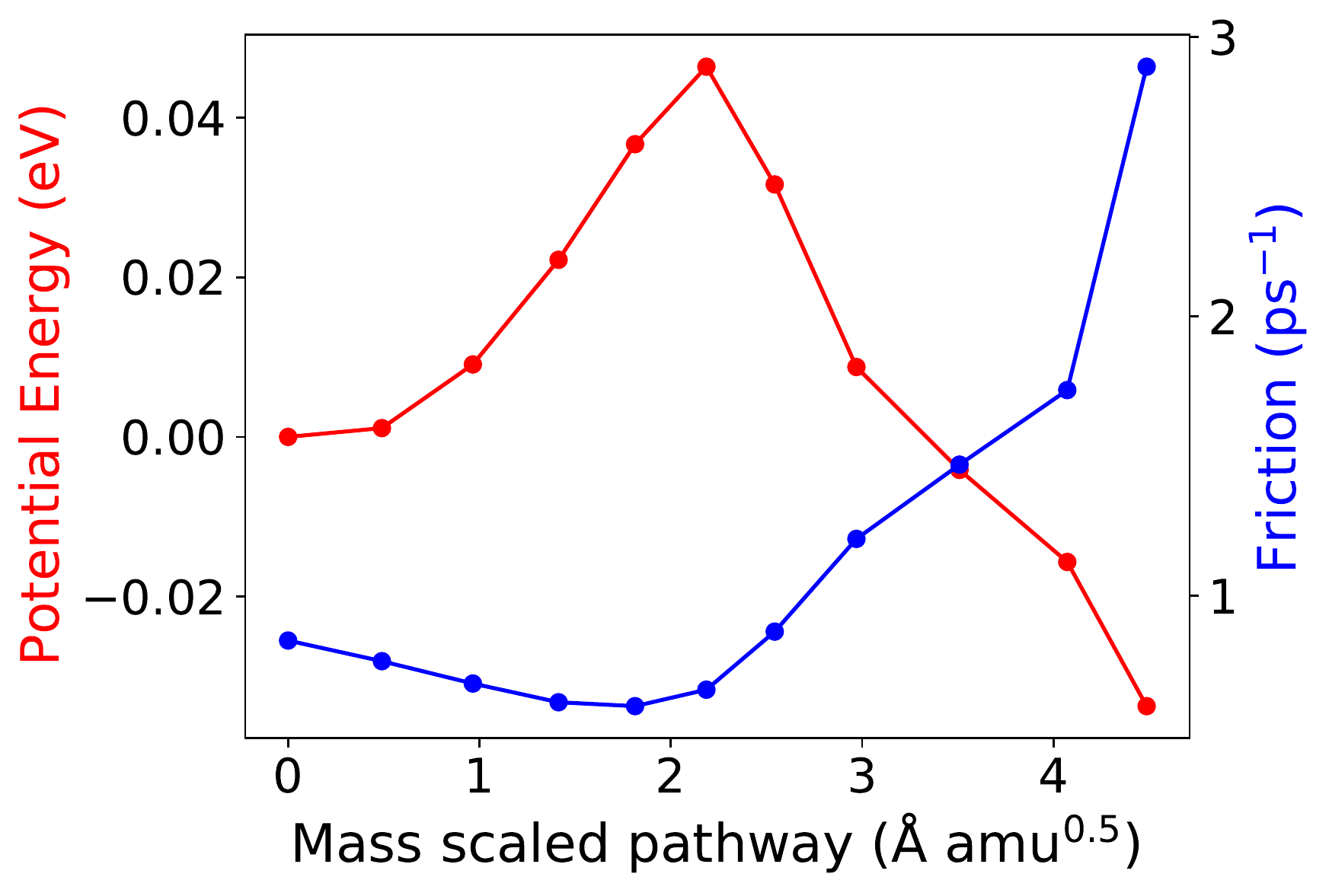}
    \caption{
     Electronic friction (Eq. 9 in the main text) and potential energy along the minimum energy pathway for Pt. The former computed only for the hydrogen atom and it is projected along the reaction coordinate. Octahedral$\rightarrow$tetrahedral transition is considered. }
    \label{fig:Pt32H}
\end{figure}

\begin{figure}[h!]
    \centering
    \includegraphics[width=0.7\columnwidth]{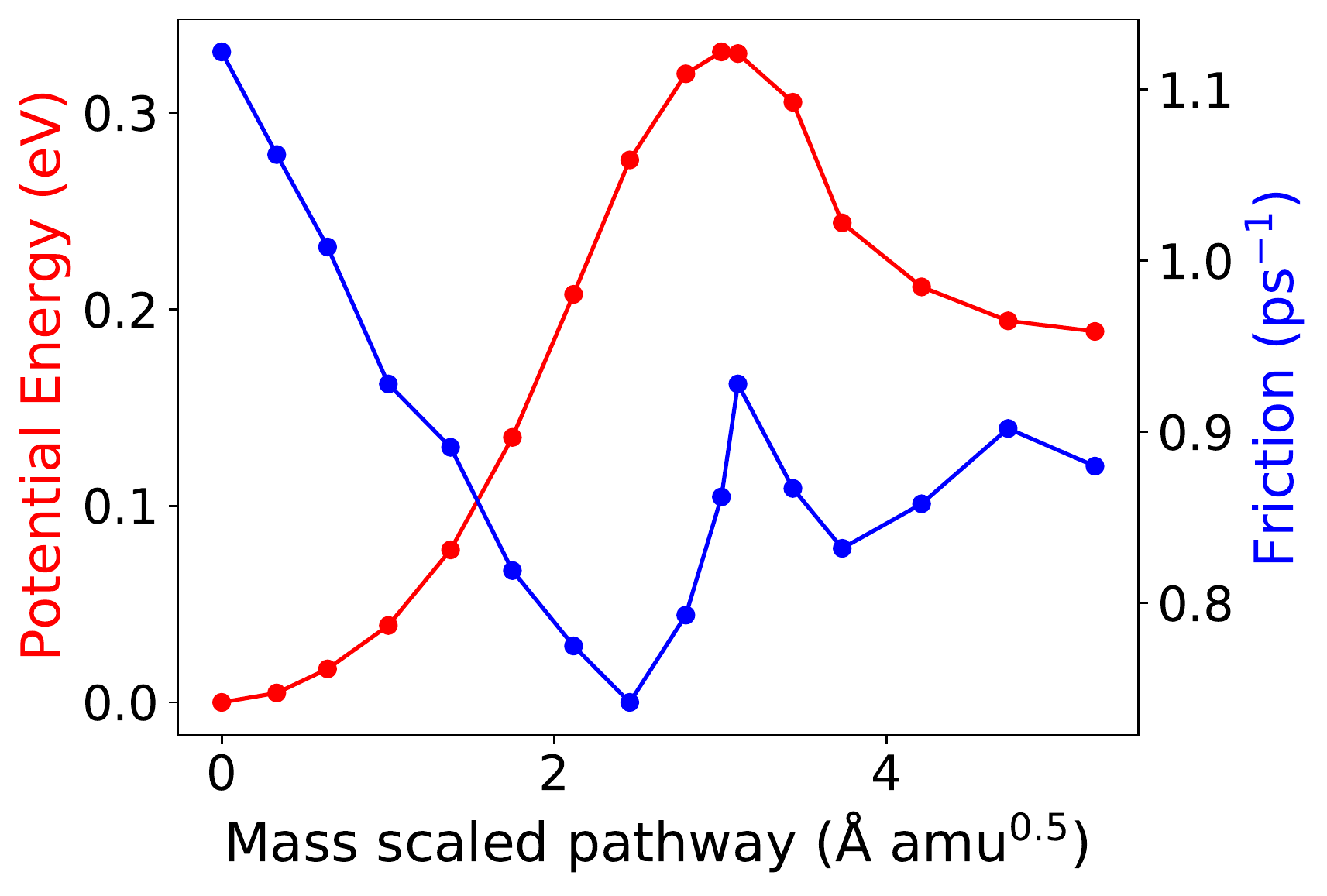}
    \caption{
    Same as \ref{fig:Pt32H} for Cu.  }
    \label{fig:Cu32H}
\end{figure}

\begin{figure}[h!]
    \centering
    \includegraphics[width=0.7\columnwidth]{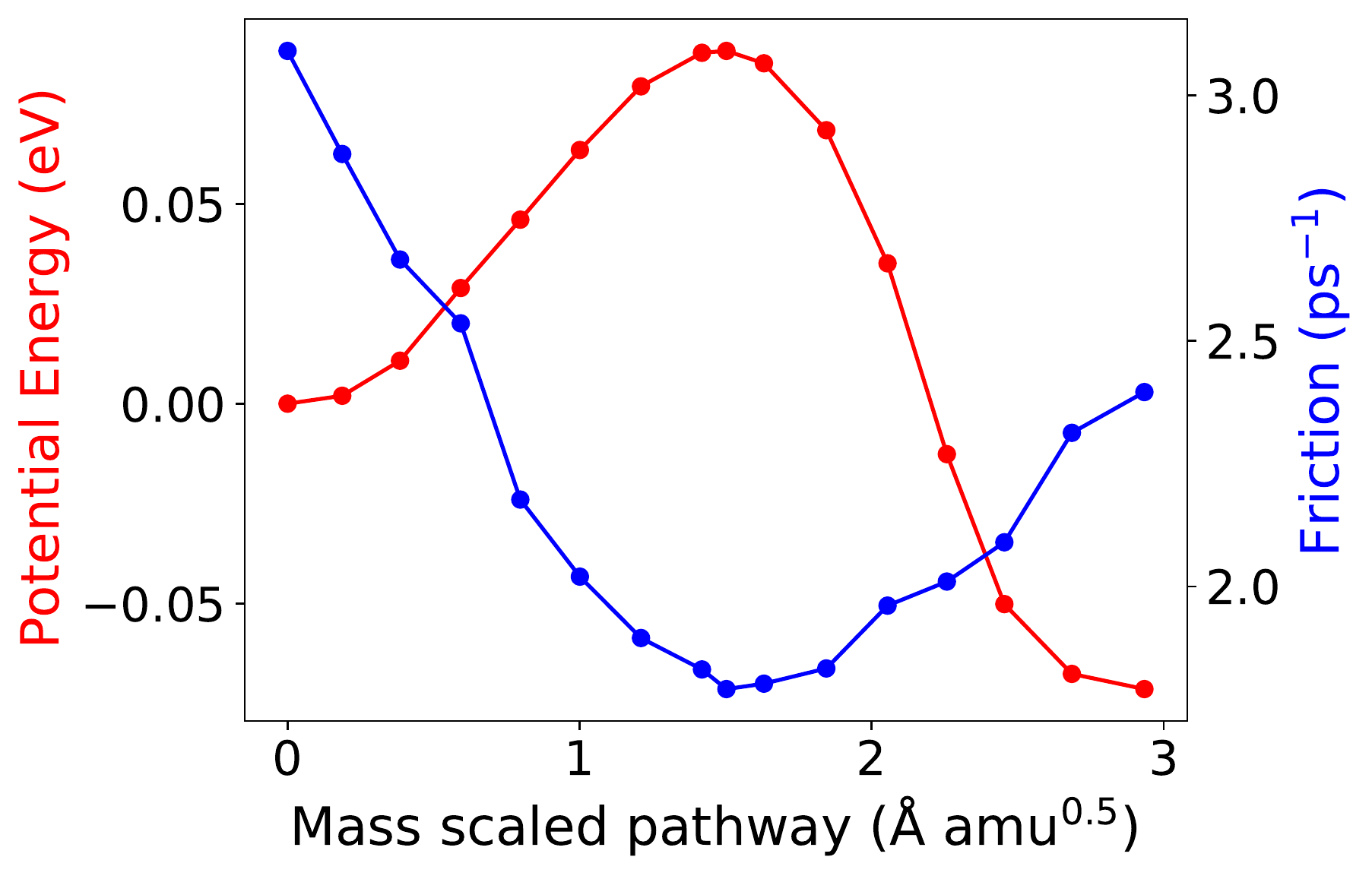}
    \caption{
     Same as \ref{fig:Pt32H} for Al.  }
    \label{fig:Al32H}
\end{figure}

\begin{figure}[h!]
    \centering
    \includegraphics[width=0.7\columnwidth]{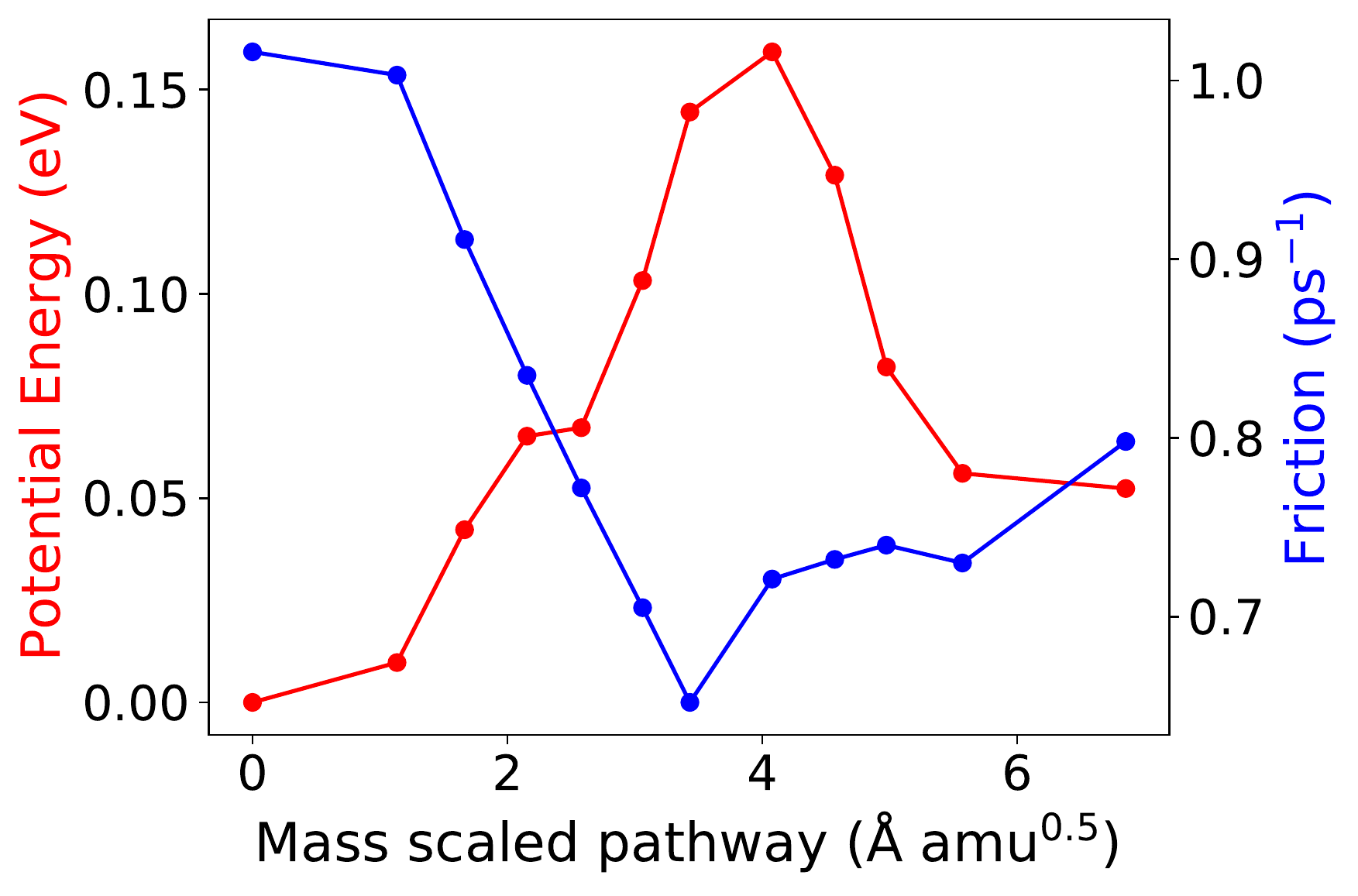}
    \caption{
     Same as \ref{fig:Pt32H} for Ag.  }
    \label{fig:Ag32H}
\end{figure}

In Fig. \ref{fig:Pt32H}-\ref{fig:Ag32H}, the friction value projected along the minimum energy path together with the minimum energy pathway (MEP) are presented.

\clearpage
\subsubsection{Frequency dependence of Electronic Friction}

\begin{figure}[h!]
     \centering
         \centering
         \includegraphics[width=0.6\textwidth]{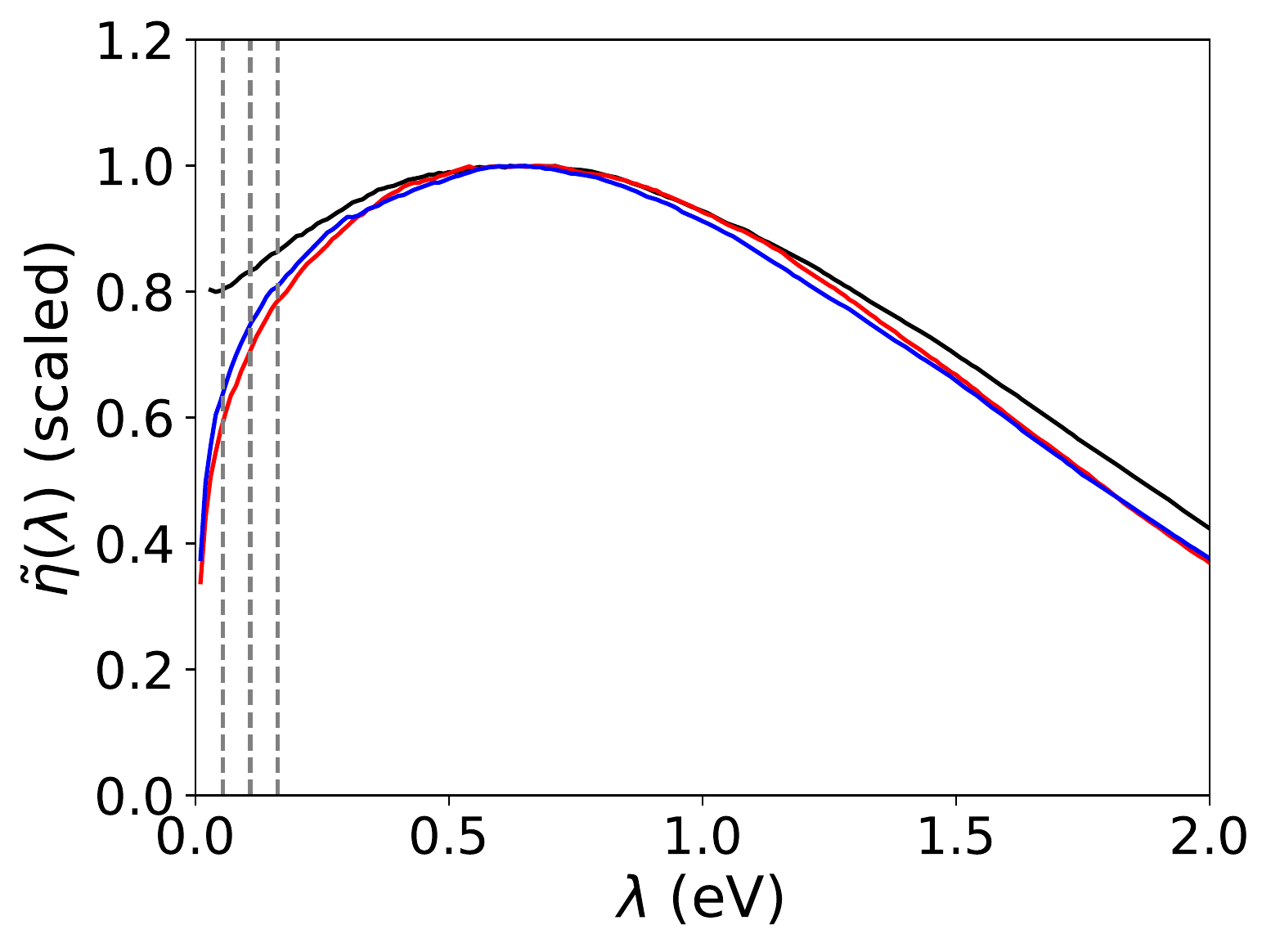}
          \caption{Frequency dependence of the friction tensor (Eq. 11 in the main text) projected on the reaction coordinate at the reactant (black), transition state (red), and product (blue) states for Pt.   The first three non-zero ring-polymer normal modes frequencies at 100K are  depicted as vertical dashed gray lines. To ease visual comparison, all curves in panel have been scaled to adopt the value of 1 at the frequency of the first local maximum. }
        \label{fig:spec_density_Pt}
\end{figure}

\begin{figure}[h!]
     \centering
         \centering
         \includegraphics[width=0.6\textwidth]{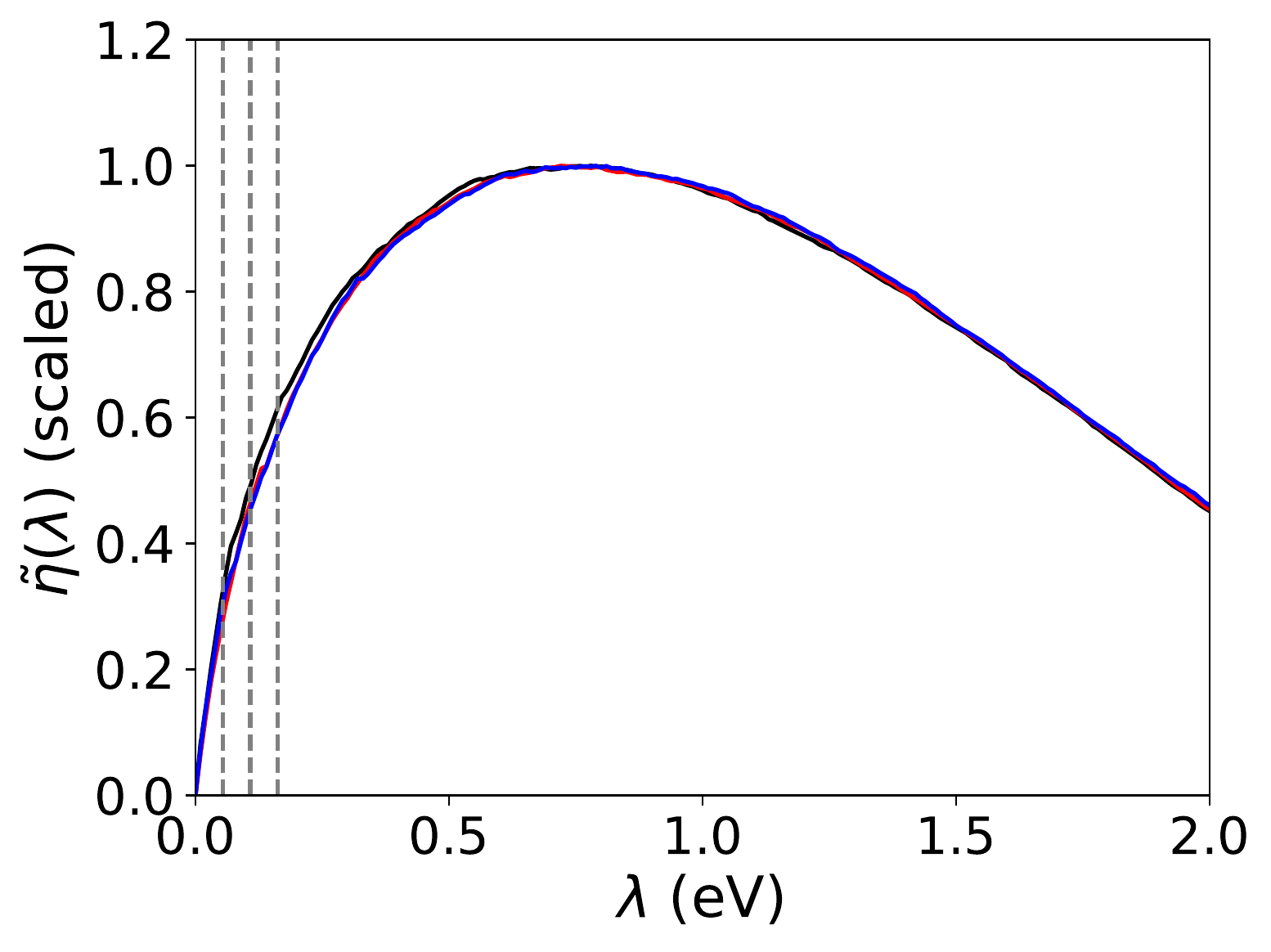}
          \caption{Same as \ref{fig:spec_density_Pt} for Al }
        \label{fig:spec_density_Al  }
\end{figure}

\begin{figure}[h!]
     \centering
         \centering
         \includegraphics[width=0.6\textwidth]{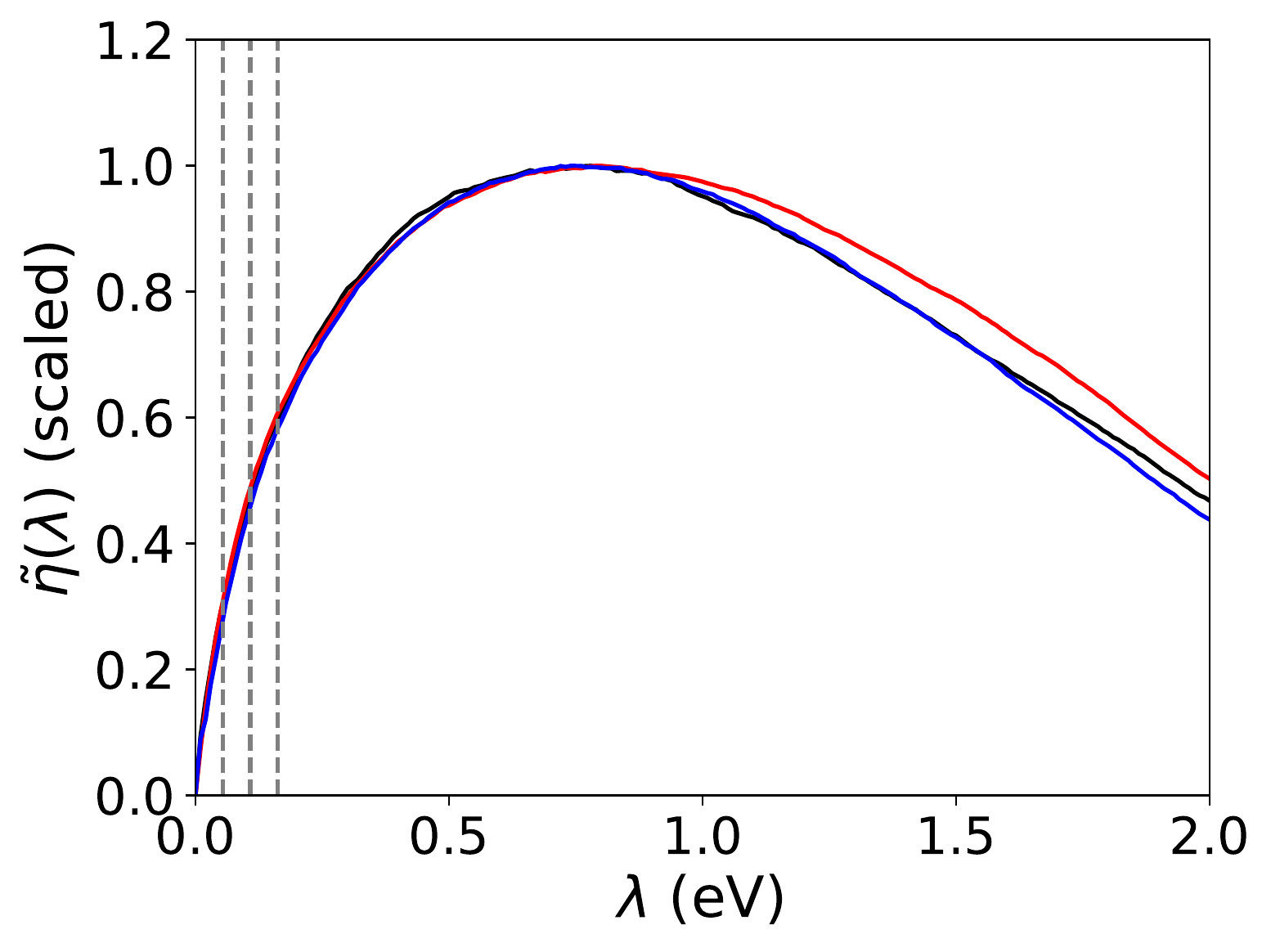}
          \caption{Same as \ref{fig:spec_density_Pt} for Cu }
        \label{fig:spec_density_Cu  }
\end{figure}

\begin{figure}[h!]
     \centering
         \centering
         \includegraphics[width=0.6\textwidth]{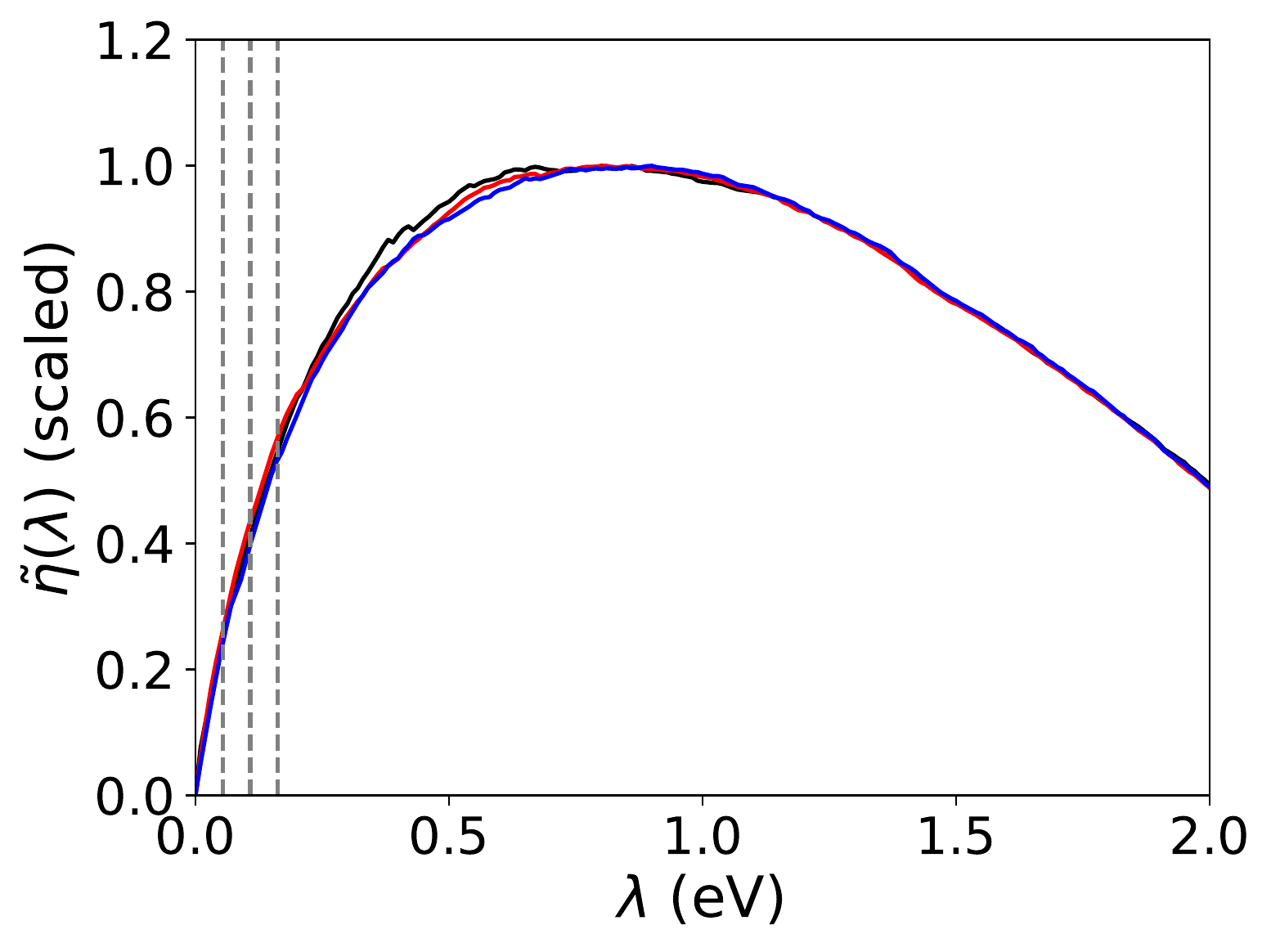}
          \caption{Same as \ref{fig:spec_density_Pt} for Ag }
        \label{fig:spec_density_Ag  }
\end{figure}

\clearpage
\subsection{Analysis of Projected Density of States }
\hfill

\begin{figure}
     \centering
     \begin{subfigure}[b]{0.4\textwidth}
         \centering
         \includegraphics[width=\textwidth]{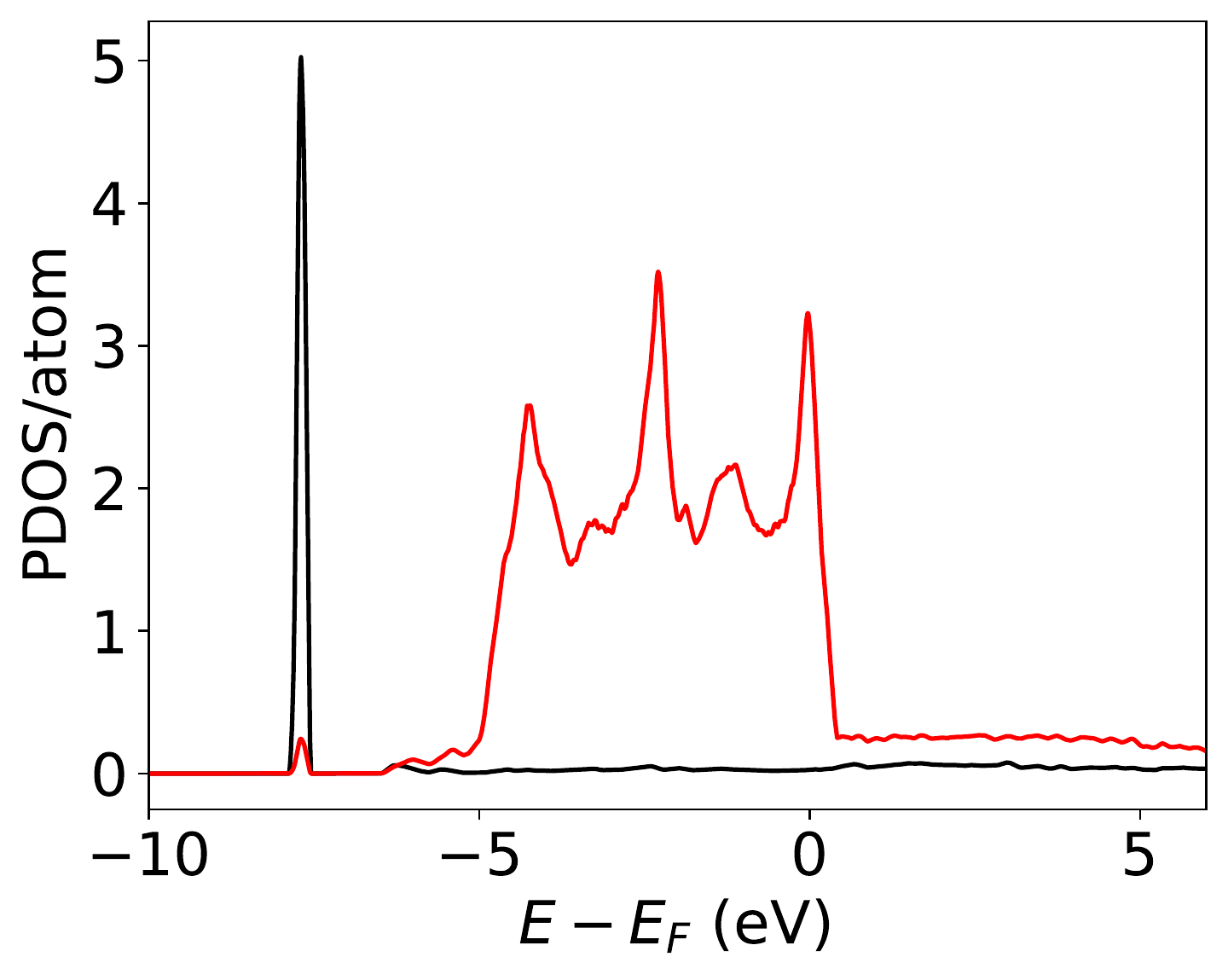}
         \caption{H@Pd reactant geometry}
         \label{fig:y equals x}
     \end{subfigure}
     \hfill
     \begin{subfigure}[b]{0.4\textwidth}
         \centering
         \includegraphics[width=\textwidth]{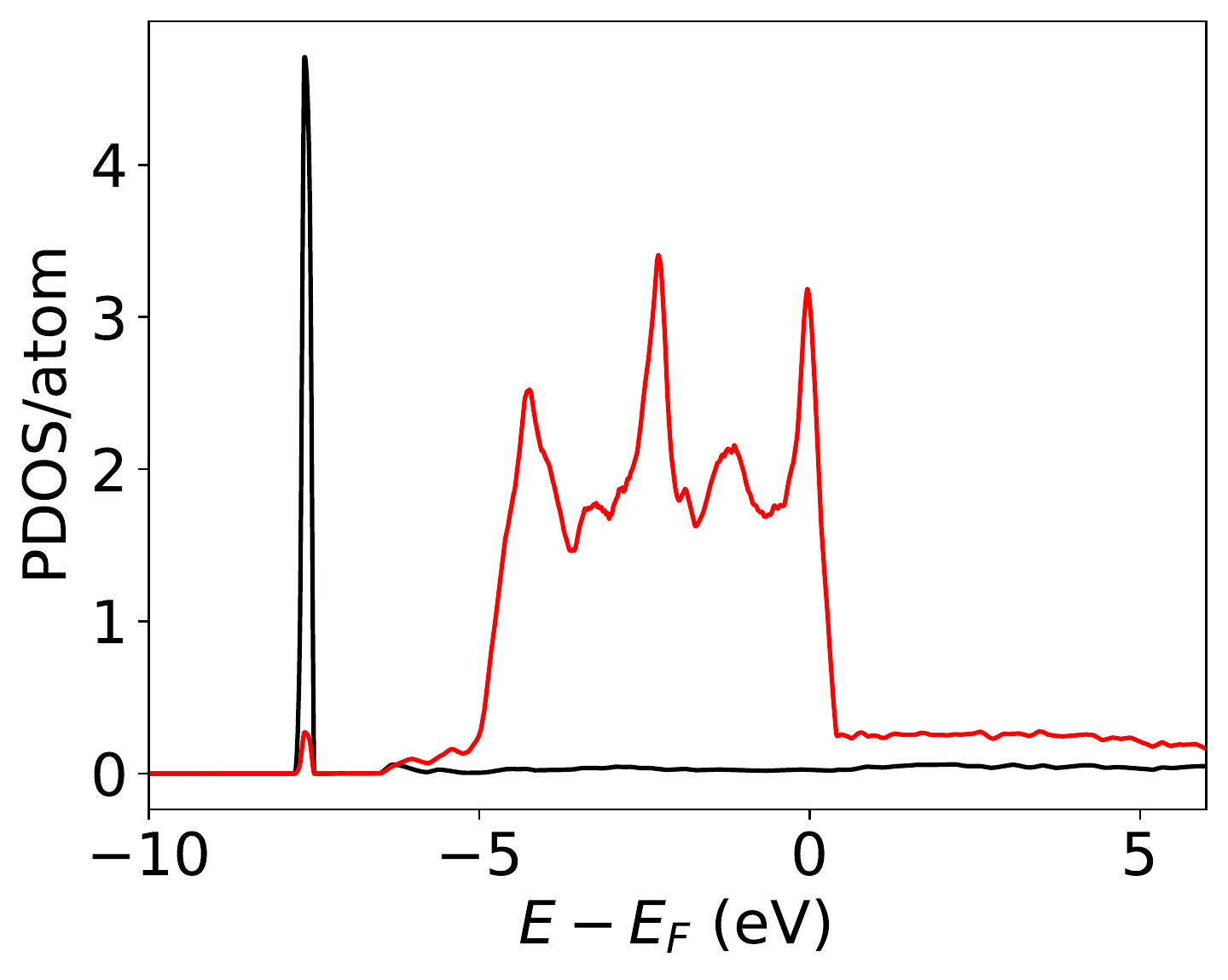}
         \caption{H@Pd transition state geometry}
         \label{fig:three sin x}
     \end{subfigure}
     \hfill
     \begin{subfigure}[b]{0.4\textwidth}
         \centering
         \includegraphics[width=\textwidth]{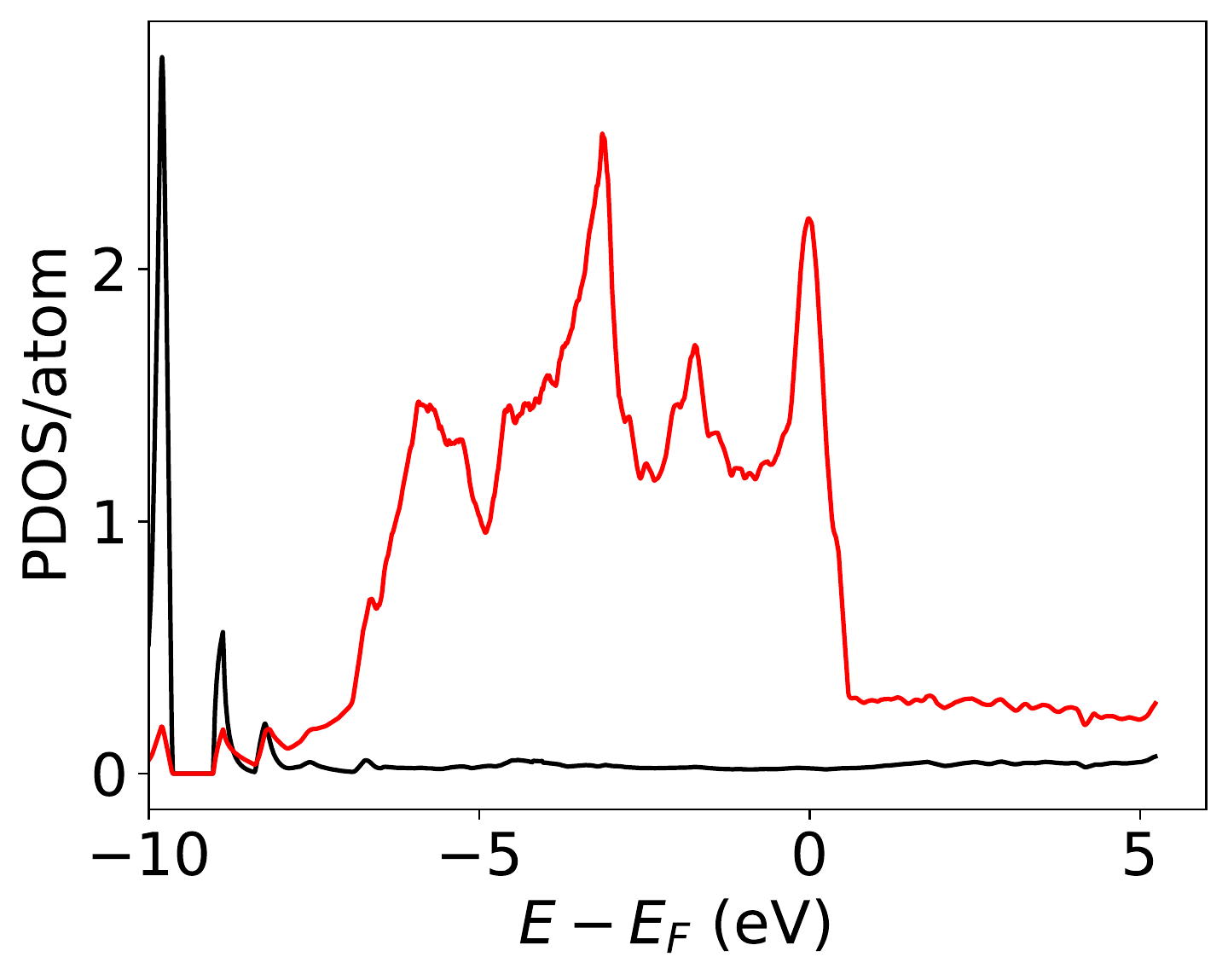}
         \caption{H@Pt reactant geometry}
         \label{fig:y equals x}
     \end{subfigure}
     \hfill
     \begin{subfigure}[b]{0.4\textwidth}
         \centering
         \includegraphics[width=\textwidth]{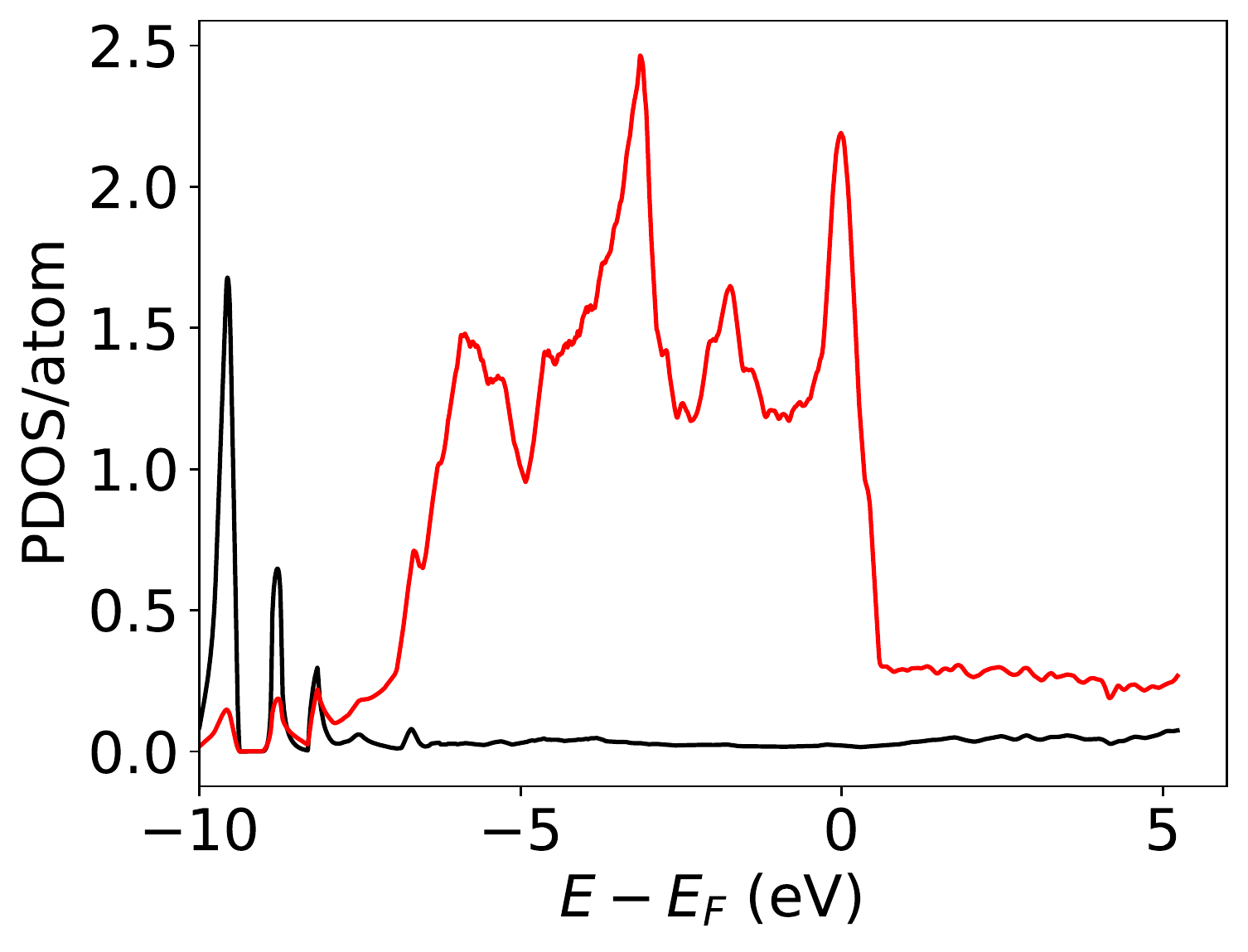}
         \caption{H@Pt transition state geometry}
         \label{fig:three sin x}
     \end{subfigure}
     \hfill
          \begin{subfigure}[b]{0.4\textwidth}
         \centering
         \includegraphics[width=\textwidth]{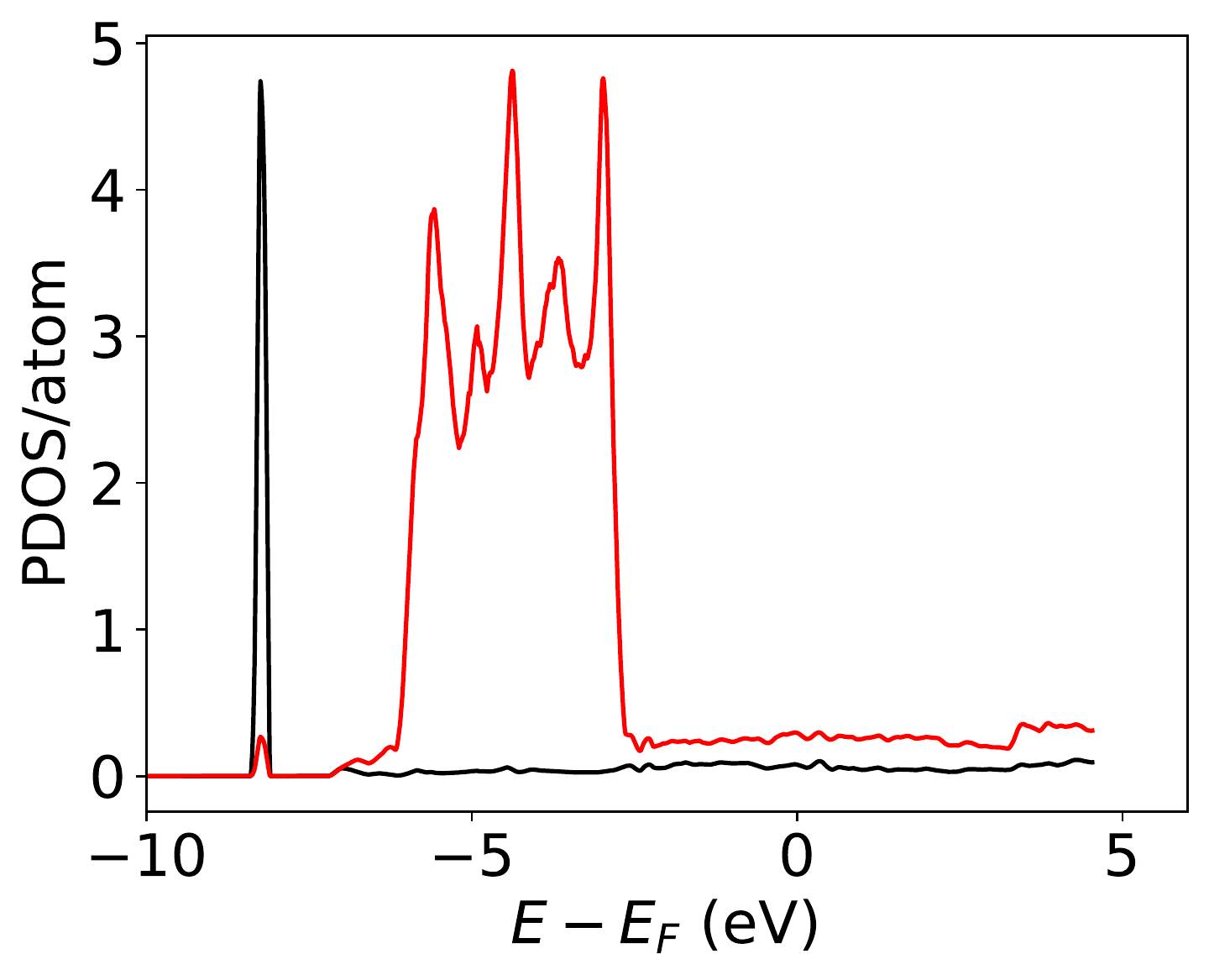}
         \caption{H@Ag reactant geometry}
         \label{fig:y equals x}
     \end{subfigure}
     \hfill
     \begin{subfigure}[b]{0.4\textwidth}
         \centering
         \includegraphics[width=\textwidth]{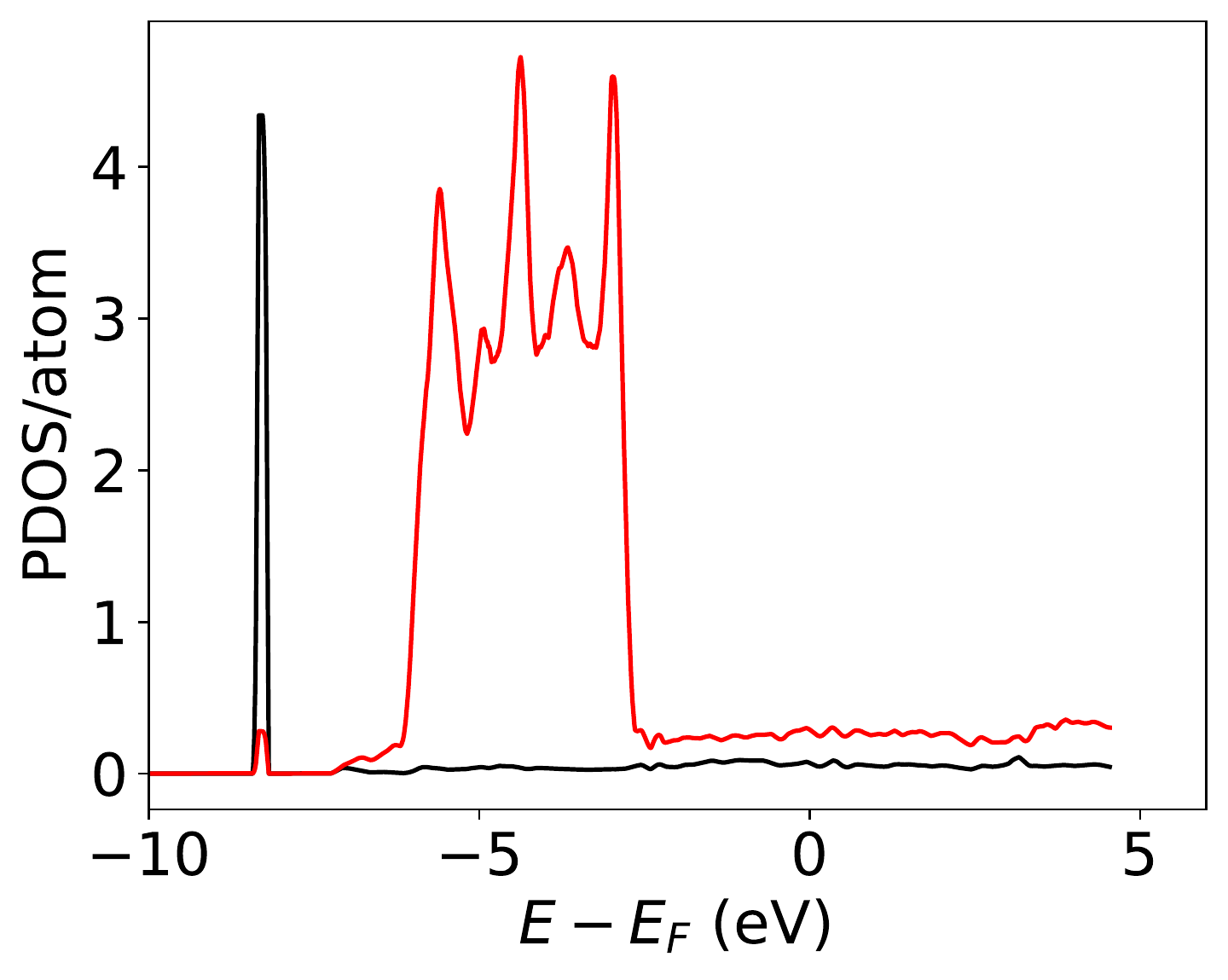}
         \caption{H@Ag transition state geometry}
         \label{fig:three sin x}
     \end{subfigure}
     \hfill
        \caption{Projected density of states (PDOS) on H (black) and X (red) atoms for X=Pd, Pt, Ag.}
        \label{fig:PDOS1}
\end{figure}

\begin{figure}
     \centering
     \begin{subfigure}[b]{0.4\textwidth}
         \centering
         \includegraphics[width=\textwidth]{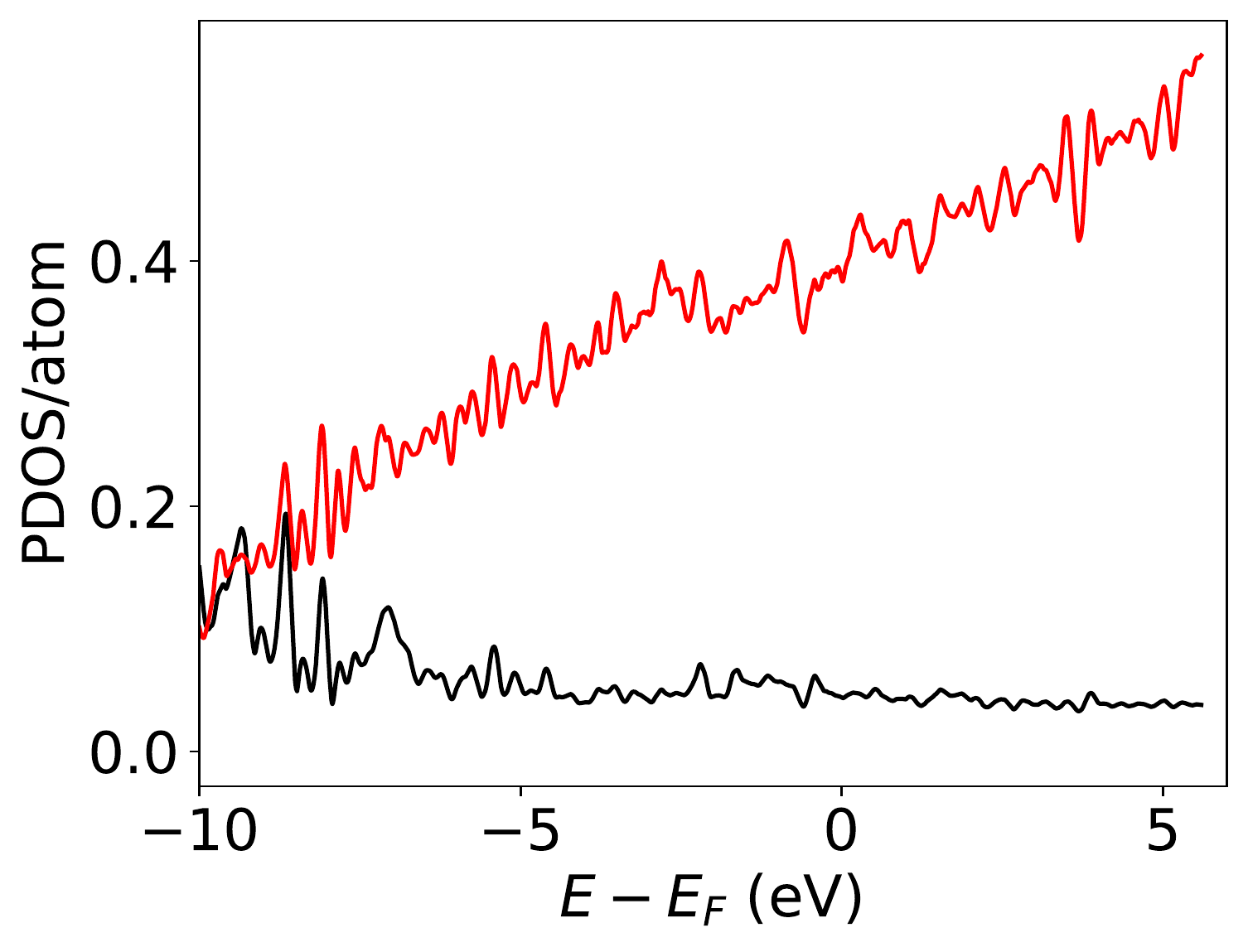}
         \caption{H@Al reactant geometry}
         \label{fig:y equals x}
     \end{subfigure}
     \hfill
     \begin{subfigure}[b]{0.4\textwidth}
         \centering
         \includegraphics[width=\textwidth]{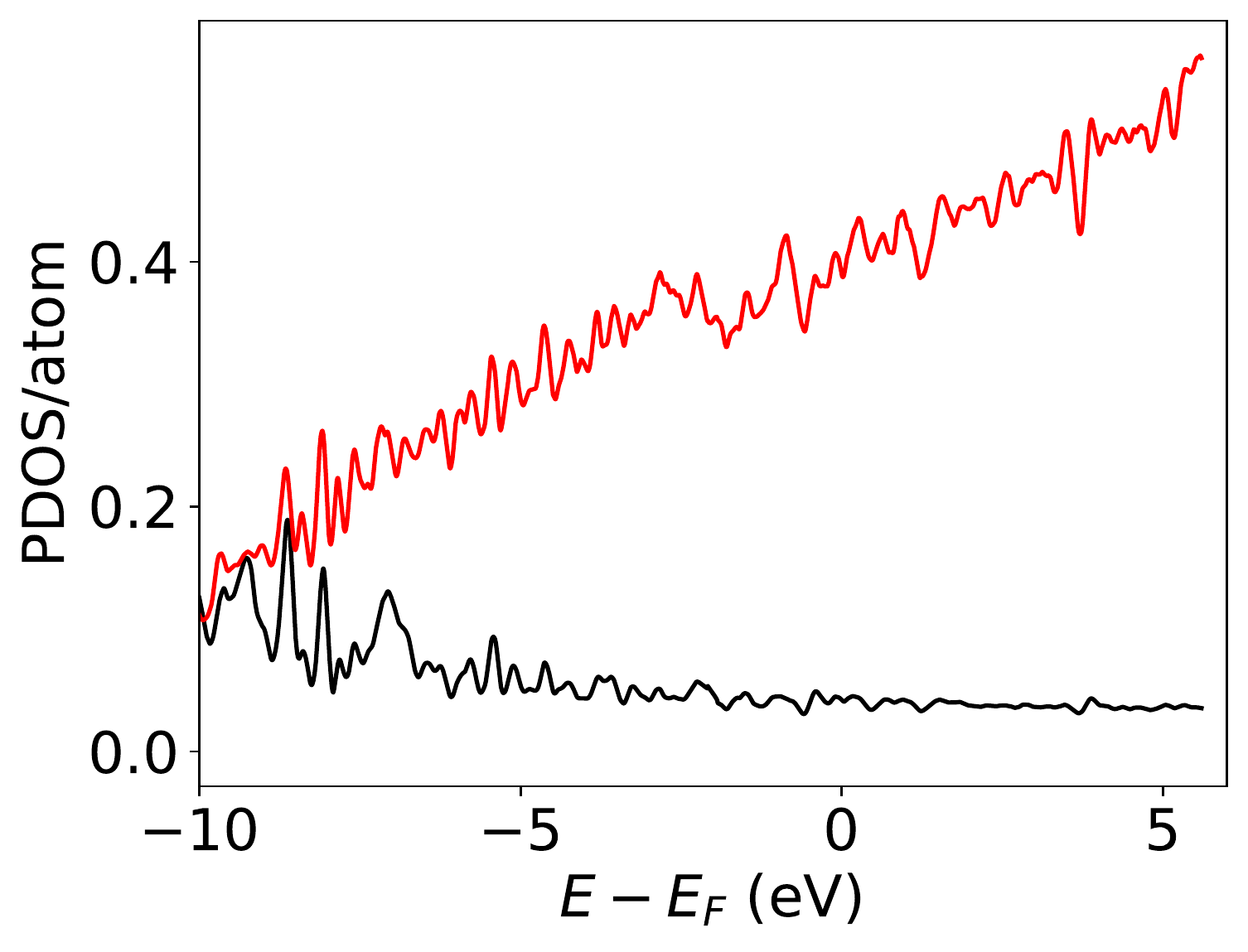}
         \caption{H@Al transition state geometry}
         \label{fig:three sin x}
     \end{subfigure}
     \hfill
          \begin{subfigure}[b]{0.4\textwidth}
         \centering
         \includegraphics[width=\textwidth]{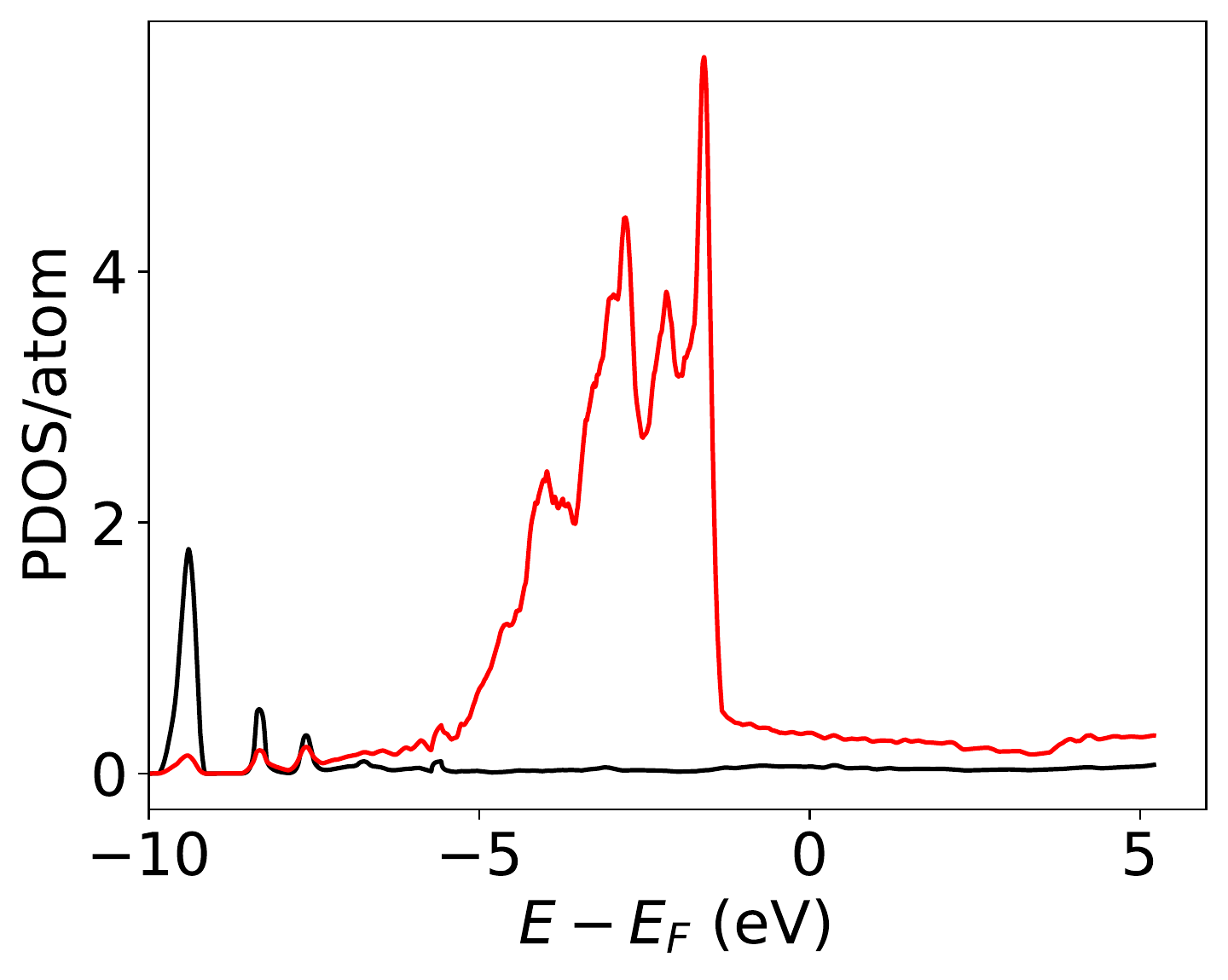}
         \caption{H@Cu reactant geometry}
         \label{fig:y equals x}
     \end{subfigure}
     \hfill
     \begin{subfigure}[b]{0.4\textwidth}
         \centering
         \includegraphics[width=\textwidth]{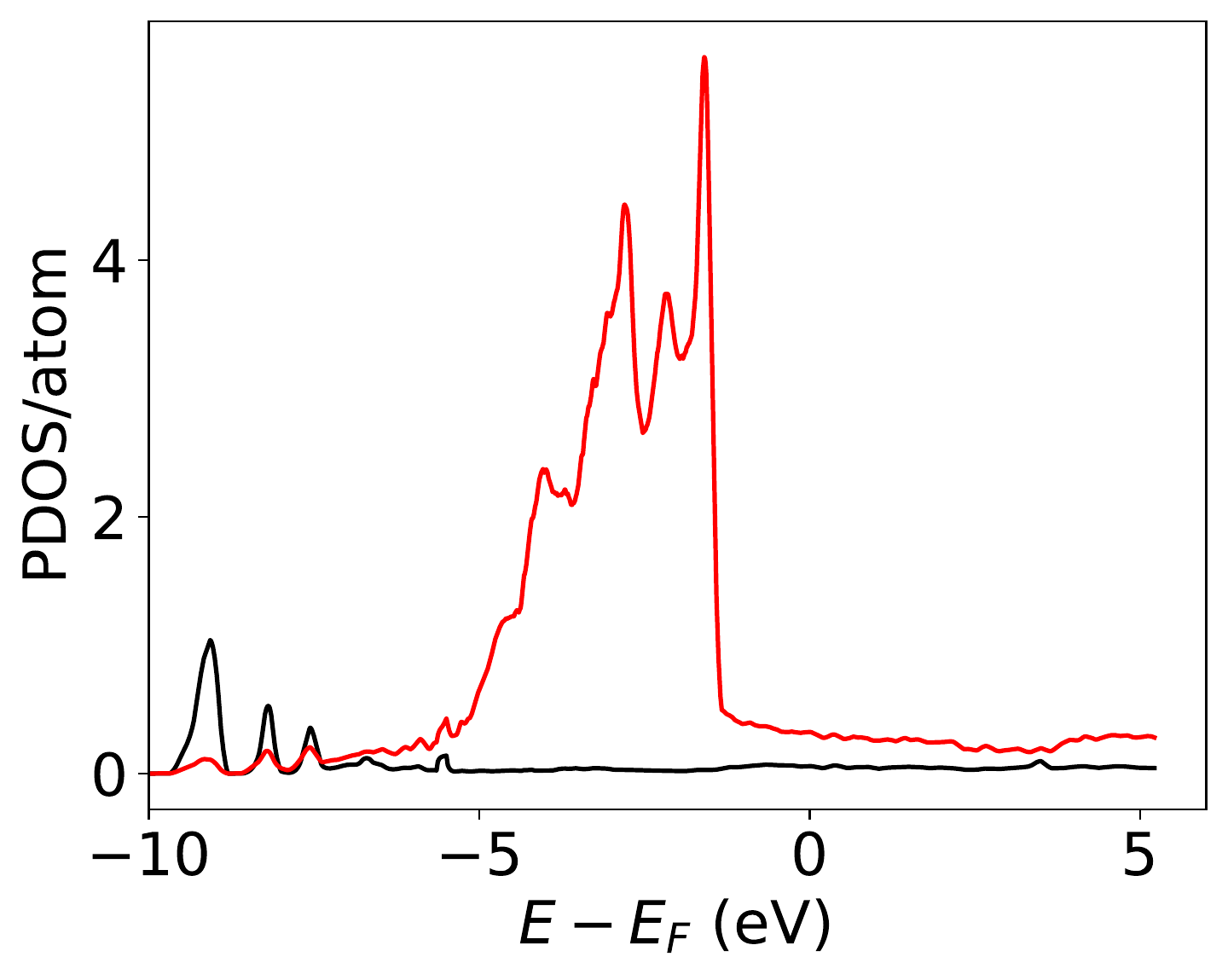}
         \caption{H@Cu transition state geometry}
         \label{fig:three sin x}
     \end{subfigure}
     \hfill
        \caption{Projected density of states (PDOS) on H (black) and X (red) atoms for X=Cu, Al}
        \label{fig:PDOS2}
\end{figure}

In Fig. \ref{fig:PDOS1} and \ref{fig:PDOS2} we show the projected density of states (PDOS) for all the systems considered in this work. These calculations were performed with a  16 $
\times$ 16 $\times$ 16 k-point sampling as used for the  friction tensor calculations.



\clearpage

\pagebreak

\section{Accuracy of Rate Estimation for Asymmetric Double Well Potential}

\begin{figure}[h!]
    \noindent
    \centering
    \includegraphics[width=.8\columnwidth]{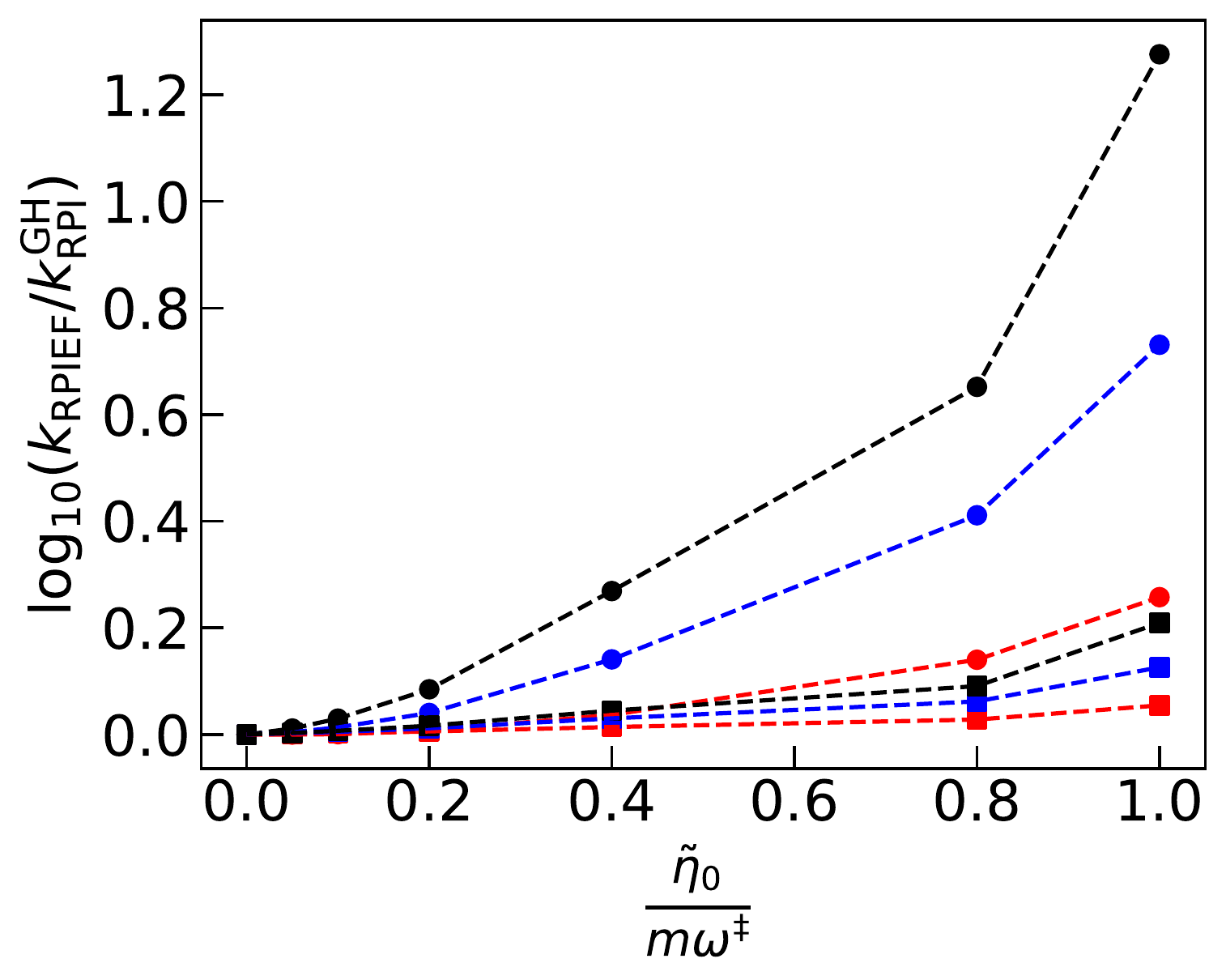}
    \caption{Error of the RPI rate values obtained by the scaling relation shown in Eq. 13 of the main text ($k_{\text{RPI}}^{\text{GH}}$), compared to the RPI-EF rates ($k_{\text{RPI-EF}}$)  for an asymmetric double well model ($q_0$=0.08    \AA) at 0.70$T^\circ_c$ (squares) and
    0.55$T^\circ_c$  (circles) for reaction barriers  of 258 meV (red), 500 meV (blue), and 1000 meV (black).
    The error is reported as the ratio of these rates.
    Analogous plot is presented in main text for a symmetric barrier model. }
    \label{fig:asymetric-error}
\end{figure}

Fig. \ref{fig:asymetric-error} shows the error incurred by the application of the extension of the Grote-Hynes type approximation into the deep tunneling regime an asymmetric double well potential. Similarly to what is observed for a symmetric double well potential in the main text, the error increases with the increase of the friction, the increase of the barrier height, and/or the decrease of the temperature. Except for the highest barrier and friction value at the lowest temperature considered, all the estimated rates are within 1 order of magnitude of the reference values. 

\bibliographystyle{aipnum4-1}

%% file: SI-MCTDH.tex
\section{MCTDH calculations}

\subsection{Flux-side approach}

The thermal rate constant $k_{\beta}$ of the isomerization reaction
$\textrm{A}\rightleftharpoons\textrm{B}$ is traditionally expressed
as the long time limit of a time-dependent ``rate constant'' $k_{\beta}(t)$,
$k_{\beta}=\lim_{t\rightarrow\infty}k_{\beta}(t)$\cite{Yamamoto1960,Miller1983,Miller1998}. The latter involves
the so-called flux-side correlation function $C_{\text{fs}}^{\beta}(t)$
and, in its refined (improved) form\cite{Drozdov2001}, reads as
\begin{equation}
k_{\beta}(t)=\frac{1}{Z_{A}}\frac{C_{\text{fs}}^{\beta}(t)}{P_{A}(0)+\left[P_{A}(0)-1\right]\chi_{\beta}-\left(\frac{1}{Z_{A}}+\frac{1}{Z_{B}}\right)\int_{0}^{t}C_{\text{fs}}^{\beta}(\tau)d\tau}.\label{eq: improved flux-side}
\end{equation}
Here, $h$ is the projection operator that separates the configuration
space of the products from that of the reagents, $Z_{A}$ and $Z_{B}$
are, respectively, the reagent and product partition functions,
\[
Z_{A}=\textrm{Tr}\left(e^{-\beta H}\left(1-h\right)\right)\ \ \ Z_{B}=\textrm{Tr}\left(e^{-\beta H}h\right)
\]
$\chi_{\beta}=Z_{A}/Z_{B}$ is the inverse equilibrium constant of
the reaction, $P_{A}(0)$ is the initial population of the reactants'
well
\[
P_{A}(0)=1-\frac{1}{Z_{A}}\textrm{Tr}\left(e^{-\beta H/2}\left(1-h\right)e^{-\beta H/2}h\right)
\]
and 
\begin{equation}
C_{\text{fs}}^{\beta}(t)=\text{Tr}\left(F_{\beta}h(t)\right)\label{eq:flux-side correlation}
\end{equation}
is the above mentioned flux-side correlation function in its most popular, symmetrized form\cite{Miller1983}. The latter
is the key quantity and requires, besides $h$ above, the Boltzmannized
flux operator
\begin{equation}
F_{\beta}=e^{-\beta H/2}Fe^{-\beta H/2}\ \ \ F=\frac{i}{\hbar}[H,h]\label{eq:Boltzmannized flux operator}
\end{equation}
which is the Boltzmannized version of the Heisenberg time derivative
of $h$ (\emph{i.e.}, the flux operator $F$). In Eq. \ref{eq: improved flux-side}
the long time limit is the true infinite time limit (\emph{i.e.},
the notation $t\rightarrow\infty$ is exact, with no \emph{caveats}),
although in practice the appropriate $t$ is a macroscopically small
time beyond which $k_{\beta}(t)$ approaches a constant value. Since
most often it holds
\[
P_{A}(0)\approx1\ \mathrm{ and} \quad \ \int_{0}^{t_{P}}C_{\text{fs}}(\tau)d\tau\ll\frac{Z_{A}Z_{B}}{Z_{A}+Z_{B}},
\]
the rate takes the form 
\begin{equation}
k_{\beta}\approx\frac{1}{Z_{A}}\lim_{"t\rightarrow\infty"}C_{\text{fs}}(t)\label{eq:flux-side}
\end{equation}
where now the limit needs to be interpreted at the \emph{plateau}
time $t_{P}$ where the correlation function attains a constant value.
Eq. \ref{eq:flux-side} represents the celebrated ``flux-side''
expression of the thermal rate constant. For a derivation of Eq. \ref{eq: improved flux-side} appropriate 
to a condensed-phase environment see Ref. \onlinecite{Craig_JCP_2007}.

\subsection{Boltzmann sampling}

We focus here on practical issues that arise when numerically evaluating
the trace expressions of Eq. \ref{eq:flux-side correlation} for a condensed-phase problem involving
many degrees of freedom \cite{Wang2006a,Craig2007}. In particular, we describe the importance sampling scheme introduced by the authors of Ref.s \onlinecite{Wang2006a,Craig2007} to turn the problem into an efficient Monte Carlo sampling of the state space. We shall first describe such 'Monte Carlo wavepacket strategy'
for computing average values of observables and, later, highlight the amendments needed to evaluate the correlation functions.

In the following we assume that $H$ takes the form $H=H_{S}+H_{\text{int}}+H_{B}$
where $H_{S}$ is the system Hamiltonian, $H_{B}$ is a sum of independent
oscillator Hamiltonians and $H_{\text{int}}$ the interaction term
between the system and the bath. Accordingly, for the state space
we have $\mathcal{H}=\mathcal{H}_{S}\otimes\mathcal{H}_{B}$, and
we make use of vectors of the form $\ket{n,N}\equiv\ket{n}\ket{N}$
where $\ket{n}$ is an arbitrary system state and $\ket{N}$ an eigenstate
of the bath, i.e., for $N=\{n_{1}n_{2}..n_{k}...\}$,
\[
H_{B}\ket{N}=E_{N}\ket{N}\ \ \mathrm{and}\quad E_{N}=\sum_{k}\hbar\omega_{k}\left(n_{k}+\frac{1}{2}\right).
\]

\subsubsection{Averages of observables}

The equilibrium average value of an operator $A$ takes the form 
\[
\braket{A}=\frac{1}{Z}\text{Tr}(e^{-\beta H}A)\equiv\frac{1}{Z}\text{Tr}(e^{-\frac{\beta}{2}H}Ae^{-\frac{\beta}{2}H})=\frac{1}{Z}\sum_{n,N}\braket{\Psi_{n,N}^{\beta}|A|\Psi_{n,N}^{\beta}}
\]
where $\ket{\Psi_{n,N}^{\beta}}=e^{-\frac{\beta}{2}H}\ket{n,N}$.
This equation is best re-written in terms of \emph{normalized} vectors
$\ket{\Phi_{n,N}^{\beta}}$, which are those directly available from
the MCTDH package upon imaginary-time evolution. Clearly, for a solution
of the Bloch equation
\[
-\frac{\partial\ket{\Psi_{\beta}}}{\partial\beta}=H\ket{\Psi}\ \ \ \ket{\Psi_{\beta=0}}=\ket{\Psi_{0}}
\]
given its squared norm $p(\beta)=\braket{\Psi_{\beta}|\Psi_{\beta}}$
and the energy expectation $\epsilon(\beta)=\braket{\Psi_{\beta}|H|\Psi_{\beta}}/p(\beta)$,
we have $\frac{\partial p}{\partial\beta}(\beta)=-2\epsilon(\beta)p(\beta)$,
hence $p\left(\beta\right)=e^{-2\int_{0}^{\beta}\epsilon(\tau)d\tau}$
and $\ket{\Psi_{\beta}}=e^{-\int_{0}^{\beta}\epsilon(\tau)d\tau}\ket{\Phi_{\beta}}$.
Thus, 
\begin{align*}
\braket{A} & =\frac{1}{Z}\sum_{n,N}p_{n,N}\left(\frac{\beta}{2}\right)\braket{\Phi_{n,N}^{\beta}|A|\Phi_{n,N}^{\beta}}\\
 & =\frac{Z_{0}}{Z}\sum_{N}\frac{e^{-\beta E_{N}}}{Z_{0}}\sum_{n}W_{n,N}(\beta)\braket{A}_{n,N}^{\beta}
\end{align*}
where $Z_{0}$ is the partition function of the uncoupled bath, $\braket{A}_{n,N}^{\beta}=\braket{\Phi_{n,N}^{\beta}|A|\Phi_{n,N}^{\beta}}$
are expectation values and
\begin{equation}
W_{n,N}(\beta)=e^{+\beta E_{N}}e^{-2\int_{0}^{\beta/2}\epsilon_{n,N}(\tau)d\tau}\label{eq:thermal weight factors}
\end{equation}
are thermal factors determined 
by the average energy  $\epsilon_{n,N}(\tau)$ of the state $\ket{\Phi_{n,N}^{\tau}}$ during
the imaginary-time evolution. The ensemble-average takes thus the
form of a Boltzmann-weighted sum of terms which can be efficiently
evaluated \emph{via} Monte Carlo sampling,
\begin{equation}
\braket{A}=\frac{Z_{0}}{Z}\sum_{n}\llangle W_{n,N}(\beta)\braket{A}_{n,N}^{\beta}\rrangle\label{eq:Monte-Carlo sampling <A>}
\end{equation}
where the double bracket denotes the Boltzmann average (and $N$ is
the corresponding random variable). This expression can be further
rewritten upon noticing that  $Z=Z_{0}\sum_{n}\llangle\ W_{n,N}(\beta)\ \rrangle$
(as it follows from Eq. \ref{eq:Monte-Carlo sampling <A>} upon setting
$A=1$), hence
\begin{equation}
\braket{A}=\frac{\sum_{n}\llangle\ W_{n,N}(\beta)\braket{A}_{n,N}^{\beta}\rrangle}{\sum_{n}\llangle\ W_{n,N}(\beta) \rrangle}\label{eq:A expectation Monte-Carlo}
\end{equation}
which requires just one sampling. 

\begin{figure}
\begin{centering}
\includegraphics[width=0.9\textwidth]{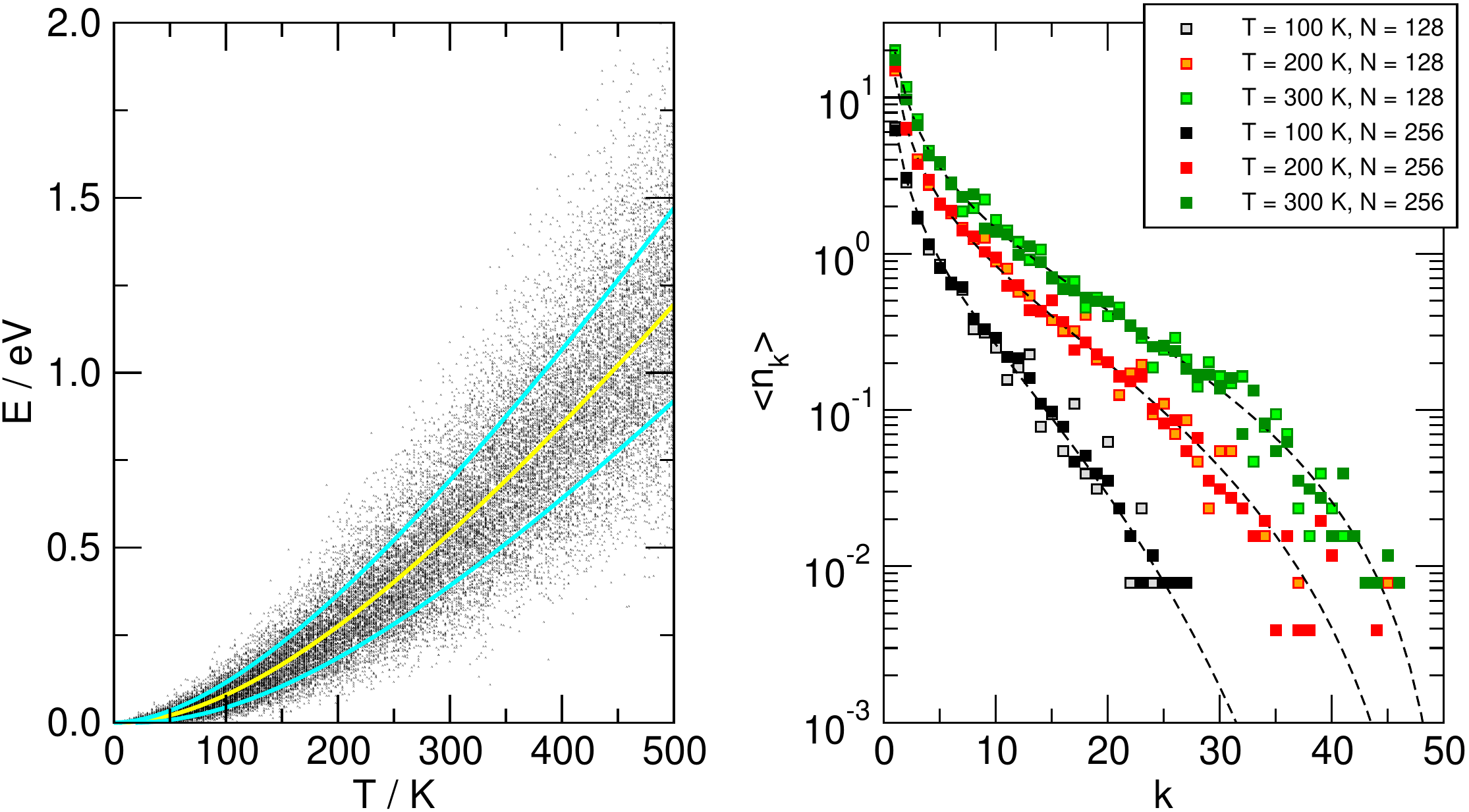}
\par\end{centering}
\caption{\label{fig:Sampling-of-the bath}Sampling of the bath (results for
a typical bath containing $F=50$ oscillators). Left: individual energy
of the bath states $\ket{N_{I}}$ sampled at different temperatures
($N_{\text{MC}}=128$ realizations for clarity), along with the theoretical
mean energy $\braket{E}$ (yellow) and $\braket{E}\pm\Delta E$ (cyan),
where $\Delta E^{2}$ is the energy variance. Right: average occupation
numbers $n_{k}$ obtained for $N_{\text{MC}}=128,256$ realization,
compared with the exact results (dashed lines).}
\end{figure}
In practice, given a basis of $\mathcal{H}_{S}$ a number of bath
configurations $\ket{N_{i}}$ are generated from the Boltzmann distribution
and the vectors $\ket{n,N_{i}}$ are propagated in imaginary-time
to compute the expectation values and the thermal factors. Sampling
can be performed independently for each bath oscillator, using the
known cumulative distributions $P_{\nu}=\sum_{k=0}^{\nu}\rho_{k}$ (where
$\rho_{k}$ are the Boltzmann probabilities) of a harmonic oscillator
with frequency $\omega$,
\[
P_{\nu}=1-e^{-\beta\hbar\omega(\nu+1)}.
\]
That is, $\nu_{i}=\text{int}(-\ln(\xi_{i}/\beta\hbar\omega))$, for
$\xi_{i}$ random in $[0,1]$, gives a set of quantum numbers $\nu_{i}$
Boltzmann distributed. Thus, besides bookkeeping issues, the procedure
is entirely straightforward and generates vectors $\ket{N_{I}}\equiv\ket{n_{i_{1}}}\ket{n_{i_{2}}}..\ket{n_{i_{k}}}..\ket{n_{i_{F}}}$
where $I=(i_{1}i_{2}..i_{k}..i_{F})$ and $F$ is the number of bath
degrees of freedom. Fig. \ref{fig:Sampling-of-the bath} illustrates
typical results of such sampling for the case considered in the main
text, which used $F=50$ harmonic oscillators with frequencies 
\[
\omega_{k}=-\omega_{c}\ln\left(\frac{k}{F+1}\right)
\]
where $\omega_{c}$ is a cutoff frequency. This is a widely used discretization
of the Ohmic bath with exponential cutoff, $J(\omega)=m\gamma\omega e^{-\omega/\omega_{c}}$,
provided the couplings $c_{k}$ with the system coordinate are set
according to 
\[
c_{k}=\omega_{k}\sqrt{\frac{2}{\pi}\frac{m\gamma\omega_{c}}{F+1}M}
\]
where $m$ and $M$ are, respectively, the system and the oscillators
mass, and $\gamma$ is the damping coefficient ($\gamma=\tilde{\eta}_0/m$, with $\tilde{\eta}_0$ being the static friction coefficient as defined in the main text.). Here, the cutoff frequency
was set to $\omega_{c}=500$ cm$^{-1}$. 

Importantly, we notice that the reagent partition function takes a form similar to the total partition function 
\begin{equation}
Z_{A}=Z_{0}\sum_{n}\llangle\ W_{n,N}(\beta)\braket{h_{A}}_{n,N}^{\beta}\ \rrangle\label{eq:reagent partition function Monte Carlo}
\end{equation}
but now the system states are better chosen to closely resemble ``the
reagent states'' since, in practice, a quickly convergent sum over
$n$ is highly desirable. In our implementation they are eigenstates
of a fictitious Hamiltonian that describes reagents only, \emph{i.e.
}the states $\ket{n}$ are chosen to be eigenstates of $\tilde{H}_{S}=\frac{p^{2}}{2m}+\tilde{v}$
where $\tilde{v}$ is a modified potential with the ``reaction channel
artificially closed''. This choice improves much the Monte Carlo
convergence and, in practice, requires 1-2 system states (depending
on the temperature) to obtain a numerically converged value of the
reagent partition function. Notice further that the convergence depends
also on other numerical parameters. For instance, a grid for the bath
oscillations that is ``centered'' around the classical equilibrium
position (the reagent minimum of the potential) is much more efficient
that an ``unbiased'' grid (like the one appropriate for the flux
evaluation). That is, for the typical Hamiltonian
\[
H=H_{S}+\sum_{k}\left[\frac{p_{k}^{2}}{2M}+\frac{M\omega_{k}^{2}}{2}\left(x_{k}-\frac{c_{k}s}{M\omega_{k}^{2}}\right)^{2}\right]
\]
the center of the grid for the $k^{\text{th}}$ harmonic oscillator
is better placed at 
\[
x^{\ominus}=\frac{c_{k}}{M\omega_{k}^{2}}s_{\text{min}}
\]
where $s_{\text{min}}$ is the value of the system coordinate at the
bottom of the reagent well. 

\subsubsection{Flux-side correlation}

\begin{figure}
\begin{centering}
\includegraphics[width=1\textwidth]{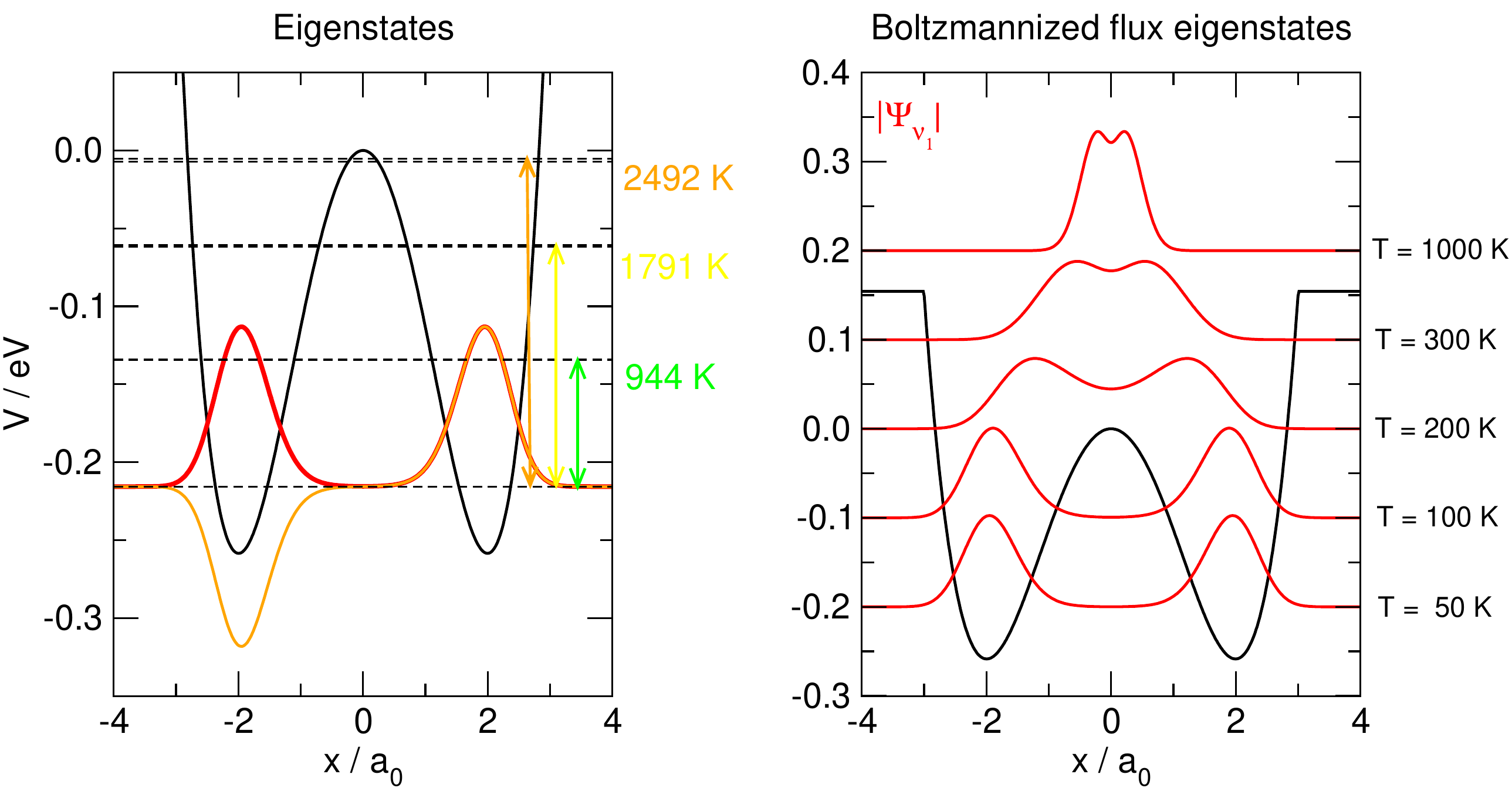}
\par\end{centering}
\caption{\label{fig:Boltmannized-flux-eigenstates}Boltmannized flux eigenstates
$\ket{\nu_{\beta}}$ for a symmetric double well potential (black
curves). Left: the two lowest-lying energy eigenstates forming the
ground-state tunneling-split doublet. Right: the absolute value of
the Boltzmannized-flux eigenfunction with the largest positive eigenvalue,
as obtained at the indicated temperatures (vertically shifted for
clarity). $a_0$ refers to the Bohr radius.}
\end{figure}

The standard flux-side correlation $C_{\text{fs}}^{\beta}\equiv\text{Tr}(e^{-\frac{\beta}{2}H}Fe^{-\frac{\beta}{2}H}h(t))$
can be evaluated similarly to the expectation values considered above
by introducing the spectral representation of $F=\sum_{u}u\ket{u}\bra{u}$
and re-writing the trace as a sum over states
\[
C_{\text{fs}}^{\beta}(t)=\sum_{u}\sum_{N}u\braket{uN|h(\tau)|uN}
\]
where the expectation of the operator $h(\text{\ensuremath{\tau}})=e^{\frac{i}{\hbar}H\tau^{*}}he^{-\frac{i}{\hbar}H\tau}$
(with $\tau=t-i\frac{\hbar\beta}{2}$) requires both the imaginary
and the real time evolution. This general procedure is, however, numerically
inconvenient since $F$ (albeit \emph{system-only}) has many contributing
eigenstates, thereby making the sum over $u$ rather long\footnote{Notice that the total number of wavepackets is given by $M\times N_{\text{MC}}$,
where $M$ is number of contributing system states and $N_{\text{MC}}$
the size of the Monte Carlo ensemble for the bath, typically $N_{\text{MC}}=128-256$.}. Fortunately, at least for the case we are interested in, there is
a way out of this ``dimensionality'' problem: the \emph{Boltzmannized}
flux operator is intrinsically of low-rank, that is it reads as
\[
F_{\beta}^{S}=e^{-\frac{\beta}{2}H_{S}}Fe^{-\frac{\beta}{2}H_{S}}=\sum_{\nu}\nu(\beta)\ket{\nu_{\beta}}\bra{\nu_{\beta}}
\]
where only few values of $\nu(\beta)$'s are significant (in fact
just two, related to each other by time inversion, if the temperature
is low enough, as it easily follows by the two-state approximation
of the dynamics)\footnote{Notice that $F$ here is a \emph{system-only} operator and that the
above expression is of low rank only in the space of system states
$\mathcal{H}_{S}$: when the same operator is considered in the whole
space $\mathcal{H}$ each eigenflux space becomes highly degenerate.}. Fig. \ref{fig:Boltmannized-flux-eigenstates} illustrates the typical
behavior of the most important flux eigenstates at varying temperatures,
and their relation to the energy eigenstates. 
 To exploit the above property, following Craig \emph{et al. } \cite{Craig_JCP_2007}, we
write
\begin{align*}
F & =e^{+\frac{\beta}{2}H_{S}}F_{\beta}^{S}e^{+\frac{\beta}{2}H_{S}}=\sum_{\nu}\nu(\beta)e^{+\frac{\beta}{2}H_{S}}\ket{\nu_{\beta}}\bra{\nu_{\beta}}e^{+\frac{\beta}{2}H_{S}}\\
 & \equiv\sum_{\nu}w_{\nu}(\beta)\ket{\bar{\nu}_{\beta}}\bra{\bar{\nu}_{\beta}}
\end{align*}
where $\ket{\bar{\nu}_{\beta}}$ are normalized, imaginary-time backward-propagated
flux eigenvectors, 
\[
\ket{\bar{\nu}_{\beta}}=\frac{1}{\sqrt{\braket{\nu_{\beta}|e^{\beta H_{S}}|\nu_{\beta}}}}e^{\beta H_{S}/2}\ket{\nu}
\]
and $w_{\nu}(\beta)=\nu(\beta)\braket{\nu_{\beta}|e^{\beta H_{S}}|\nu_{\beta}}$
are flux-weights. The $\ket{\bar{\nu}_{\beta}}$'s are non-orthogonal, yet rather useful since they provide a compact representation of the flux operator. With these definitions, we have
\begin{align*}
C_{\text{fs}}^{\beta} & =\sum_{\nu}\text{Tr}\left(e^{-\beta H/2}\ket{\bar{\nu}_{\beta}}\bra{\bar{\nu}_{\beta}}e^{-\beta H/2}h(t)\right)w_{\nu}(\beta)\\
 & \equiv\sum_{\nu}\sum_{N}w_{\nu}(\beta)\braket{\bar{\nu}_{\beta}N|e^{-\beta H/2}h(t)e^{-\beta H/2}|\bar{\nu}_{\beta}N}
\end{align*}
and everything proceeds as above. Specifically, introducing the normalized
vectors
\[
\ket{\Phi_{\nu,N}^{\beta}}=\frac{1}{\sqrt{p_{\nu N}(\beta)}}e^{-\beta H/2}\ket{\bar{\nu}_{S}N}\ , \quad\ p_{\nu N}(\beta)=e^{-2\int_{0}^{\beta/2}\epsilon_{\nu N}(\tau)d\tau},
\]
their real-time evolutions 
\[
\ket{\Phi_{\nu,N}^{\beta}(t)}=e^{-\frac{i}{\hbar}Ht}\ket{\Phi_{\nu,N}^{\beta}}
\]
and the expectation $\braket{h(t)}_{\nu N}^{\beta}=\braket{\Phi_{\nu,N}^{\beta}(t)|h|\Phi_{\nu,N}^{\beta}(t)}$
we write
\[
C_{\text{fs}}^{\beta}=Z_{0}\sum_{N}\frac{e^{-\beta E_{N}}}{Z_{0}}\sum_{\nu}w_{\nu}(\beta)W_{\nu N}(\beta)\braket{h(t)}_{\nu N}^{\beta},
\]
where $W_{\nu N}(\beta)=p_{\nu N}(\beta)e^{\beta E_{N}}$ are thermal
factors (Eq. \ref{eq:thermal weight factors}) and $w_{\nu}(\beta)$
are (thermal) flux weights. Equivalently, 
\begin{equation}
C_{\text{fs}}(t)=Z_{0}\sum_{\nu}\llangle\ w_{\nu}(\beta)W_{\nu N}(\beta)\braket{h(t)}_{\nu N}^{\beta}\rrangle\label{eq:flux-side correlation Monte Carlo}
\end{equation}
is the working expression involving 
\begin{itemize}
\item [{\emph{(i-a)}}] a ``system'' preparation to define the appropriate
system states $\ket{\bar{\nu}_{\beta}}$, 
\item [{\emph{(i-b)}}] a Monte Carlo sampling of the (uncoupled) bath state,
which delivers bath states $\ket{N}$, 
\item [{\emph{(ii)}}] a relaxation dynamics with the full Hamiltonian
$H$ for each state $\ket{\bar{\nu}_{\beta}N}$ 
\item [{\emph{(iii)}}] a real time dynamics with the same Hamiltonian
on the relaxed states obtained in \emph{(ii)}. 
\end{itemize}
When combined with the calculation of the reagent partition function,
Eq. \ref{eq:reagent partition function Monte Carlo}, this recipe
gives the thermal rate constant of Eq. \ref{eq:flux-side} or, with
minor additional effort, the one defined by Eq. \ref{eq: improved flux-side}.

In closing this section it is worth noticing that the manipulations
above with the Boltzmann operator are well defined when the Hamiltonian
operator $H_{S}$ is bound\footnote{When a maximum eigenvalue exists the Boltzmann operator is non-singular.}
but some regularization is needed when --- as it is often the case in
realistic problems --- the spectrum of $H_{S}$ is unbound. This is particularly important for numerical applications because
the presence of high-lying energy eigenstates makes $\exp\left(-\beta H_{S}\right)$
nearly singular and $\exp\left(+\beta H_{S}\right)$ numerically unstable,
increasingly so when reducing the temperature ($\beta\rightarrow\infty$).
Fortunately, such high-lying energy eigenstates should not play any
role in the dynamics (particularly at the low temperatures where a
quantum description is required) and the simple regularization of
replacing $H_{S}$ with its projection $\bar{H}_{S}$ on a low-lying
energy eigenspace suffices. Specifically, defining $P_{n}$ the projection
onto the first (lowest lying) $n$ eigenstates of $H_{S}$ one defines
$\bar{H}_{S}=P_{n}H_{S}P_{n}$ and uses it in place of $H_{S}$ in
some of the expressions above. This is entirely legitimate since the
aim is just to re-write $F$ in a suitable way, but of course one
must ensure that $\bar{H}_{S}$ closely resembles $H_{S}$ for the
expansion to be compact. In practice, then, one uses $H_{S}$ to define
the Boltzmannized flux and diagonalize it, and then switches to $\bar{H}_{S}$
when propagating backward in imaginary time, making sure that the
``error'' is kept below a desired threshold\footnote{The reliability of such regularization can be measured by the difference
in the (differently) Boltzmannized flux eigenvalues and by two kinds
of errors in the flux eigenstates: there exists a \emph{truncation}
error (the distance between the eigenvectors of the differently Boltzmannized
operators) and a \emph{propagation} error (defined by back-propagating
with the ``projected version'' of the Boltzmann operator followed
by propagation with the true Boltzmann operator). The latter two behave
oppositely when varying $n$ since the truncation error determines
the quality of the projected Hamiltonian and the propagation error
reflects the above mentioned numerical instability.}$^{,}$\footnote{There is no need for $H_{S}$ to be the same ``system'' Hamiltonian
appearing in the total Hamiltonian, and one can set it at his own
convenience. After all the flux operator $F$ does \emph{not} depend
on the potential. }. 

\subsection{ML-MCTDH wavefunction and calculation setup}

\begin{figure}
\begin{centering}
\includegraphics[angle=90,height=0.8\textheight]{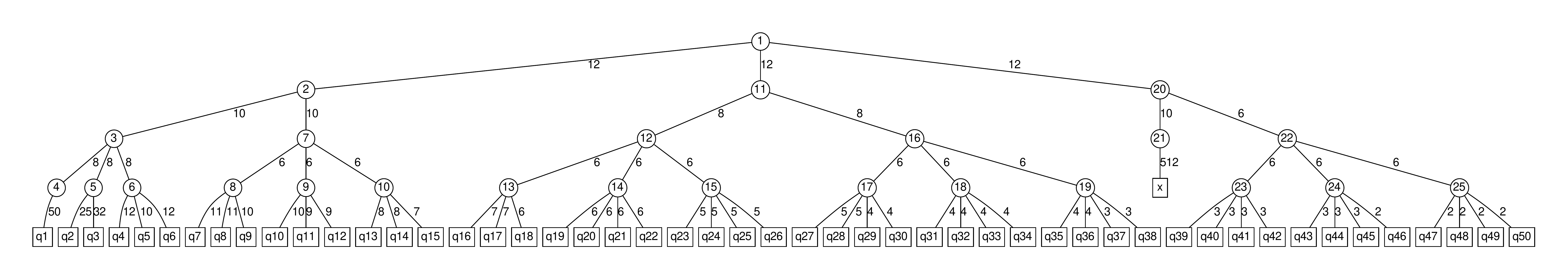}
\par\end{centering}
\caption{\label{fig:ML-tree}Adopted tree structure for the ML-MCTDH wavefunction. Each circle denotes a MCTDH-like expansion in the tree, along with its modes (the arms) and the number of the single particle functions used in the expansion (the numbers on the arms). Squares denote the bottom layers where modes reduce to physical degrees of freedom and the numbers represent the size of the primitive grids. Here $x$ is the system coordinate and the $q_k$'s are bath coordinates.}
\end{figure}
Fig. \ref{fig:ML-tree} shows the tree structure defining the ML-MCTDH wavefunction.
It was obtained after extensive testing, and found to accurately reproduce
previous ML-MCTDH and PI results, at both high and low temperatures at varying coupling strengths, ranging from the weak to the strong coupling limits. The number of logical modes
used in each layer was kept small ($2-3$) and each of them was described
with several single-particle functions ($12-6$). Only in the bottom
layer, depending on the size of the primitive grids, the bath degrees
of freedom (the ``$q_{k}$''s in Fig. \ref{fig:ML-tree}, sorted
in order of increasing frequency) were grouped in $3-4$ dimensional
single-particles. The system mode (the ``$x$'' in Fig. \ref{fig:ML-tree})
is located in the group of bath modes with comparable frequency, although
described separately from the bath modes and with the help of a large
number of SPFs. As for the primitive grids we used Harmonic Oscillator - Discrete Value Representation (HO-DVR) grids for
each bath degrees of freedom, which amounts to introduce a phonon
basis of the same size for the corresponding mode. The grids for the low
frequency modes used several tens of points (as indicated in Fig.
\ref{fig:ML-tree}), and were extended if necessary to accommodate each realization sampled from the canonical ensemble of the bath. The grids for the high-frequency
modes, on the other hand, used much fewer grid points since they were barely excited during
the dynamics. The system degree of freedom was described with a uniformly
spaced grid, a Fast Fourier Transform - Discrete Value Representation (FFT-DVR), using 512 grid points in the range $x\in[-4,4]\ a_{0}$ (with $a_0$ referring to the Bohr radius). 

\begin{figure}
\begin{centering}
\includegraphics[width=0.95\columnwidth]{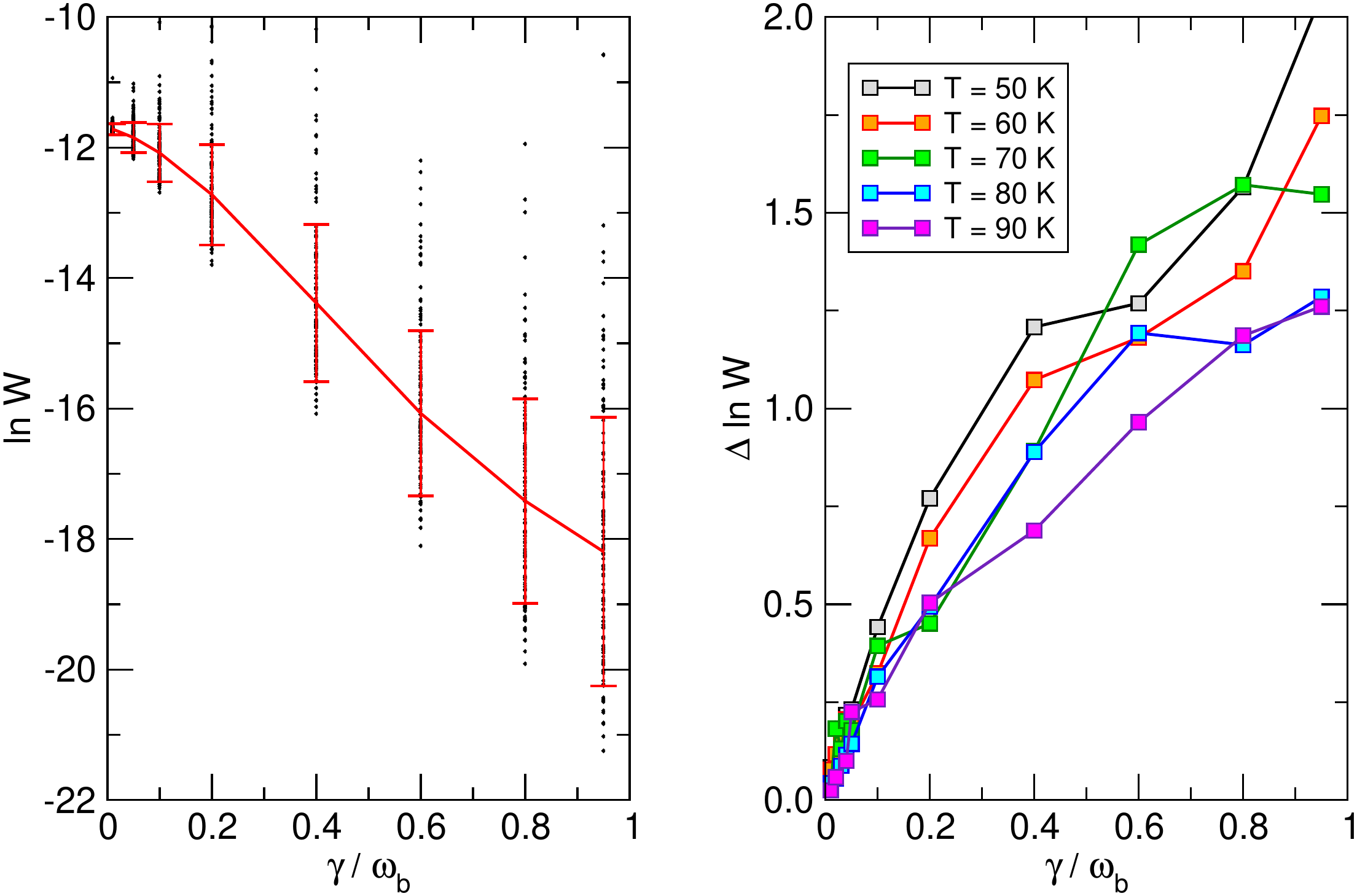}
\par\end{centering}
\caption{\label{fig:Thermal factors}Behavior of the thermal factors otained
in thermalization step. Left: individual realizations, with their
average and root mean square given as red curve with error bars. Right:
root mean square for different temperatures, as a function of the
coupling strength.}

\end{figure}
\begin{figure}
\begin{centering}
\includegraphics[width=0.95\columnwidth]{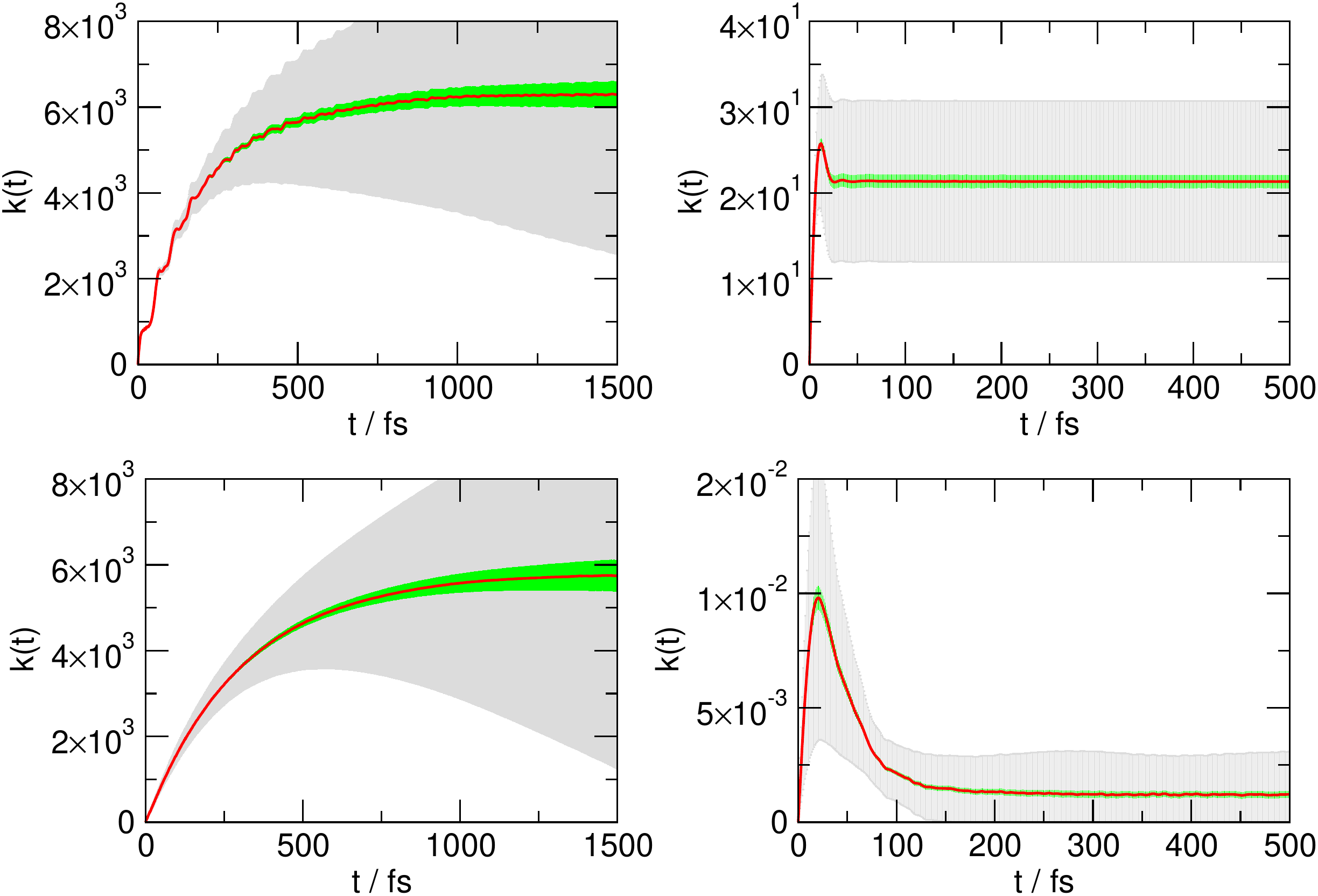}
\par\end{centering}
\caption{\label{fig:Illustrative-examples-of-Cfs}Illustrative examples showing
the behavior of the flux-side correlation function  at $T = 90$ K (top row) and $T=50$ K (bottom row),
for very small and large coupling strength, $\gamma/\omega_b=0.01$ and $0.95$, respectively for the left and right panels. Gray (green) area shows the spread (statistical error) calculated as the root-mean-square deviation of $x_N$ (root-mean-square deviation of $x_N$ over square-root of number of realizations).}

\end{figure}
The MCTDH equations of motion were integrated with the Variable Mean
Field scheme, using a variable step-size $8^{\text{th}}$-order Runge-Kutta
integrator for both the amplitude coefficients and the single-particle
wavefunctions, and a small accuracy parameter ($10^{-8}$). The propagation
time was set differently according to the coupling strength, in order
to guarantee that the computed flux-side correlation functions attained
a constant limiting value (which occurs at increasingly longer times
when decreasing friction). Similarly, the number of realization was
chosen differently depending on the considered temperature and coupling
strength, the higher $T$ and/or $\gamma$ the larger the number of
realizations used. Fig. \ref{fig:Thermal factors} shows the behavior
of the thermal factors of Eq. \ref{eq:thermal weight factors} as
computed in the ``equilibration step'' (step ii above), which are
the weights with which the individual realizations of the bath enter
into the flux-side correlation function expression of Eq. \ref{eq:flux-side correlation Monte Carlo}
(for a given flux state). Their logarithms are within $0-2$ over
a wide range of coupling strengths, meaning roughly that each realization
has a weight in a two-order-of-magnitude wide interval at most. Fig. \ref{fig:Illustrative-examples-of-Cfs},
on the other hand, displays some illustrative examples of the evolution
of the correlation functions, along with the spread of the contributing
terms in the sum of Eq. \ref{eq:flux-side correlation Monte Carlo}. More specifically, the latter was obtained by grouping results from opposite flux eigenvalues, and rewriting Eq. \ref{eq:flux-side correlation Monte Carlo} as an average of individual contributions $x_N =\sum_\nu W_{\nu N} |w_\nu (\beta)| \left[\braket{h}_{+|\nu|, N}-\braket{h}_{-|\nu|, N}\right]$ with a spread given by the root-mean-square deviation of the  $x_N$'s. Hence, provided bath sampling is sufficiently extended, it measures the intrinsic variability of the reaction probability across the thermal equilibrium state.